\documentclass[pdflatex,sn-nature]{sn-jnl}

\usepackage{graphicx}%
\usepackage{multirow}%
\usepackage{amsmath,amssymb,amsfonts}%
\usepackage{amsthm}%
\usepackage{mathrsfs}%
\usepackage[title]{appendix}%
\usepackage{xcolor}%
\usepackage{textcomp}%
\usepackage{manyfoot}%
\usepackage{booktabs}%
\usepackage{algorithm}%
\usepackage{algorithmicx}%
\usepackage{algpseudocode}%
\usepackage{listings}%
\usepackage{xcolor}
\graphicspath{{}}
\newlength{\boxwidth}
\def\btheorem{\begin{theorem}}
\def\etheorem{\end{theorem}}
\def\blemma{\begin{lemma}}
\def\elemma{\end{lemma}}
\def\bproposition{\begin{proposition}}
\def\eproposition{\end{proposition}}
\def\bcorollary{\begin{corollary}}
\def\ecorollary{\end{corollary}}
\def\bdefinition{\begin{definition}}
\def\edefinition{\end{definition}}
\def\bexample{\begin{example}}
\def\eexample{\end{example}}
\def\bremark{\begin{remark}}
\def\eremark{\end{remark}}

\newcommand{\ba}{\begin{array}}
\newcommand{\ea}{\end{array}}

\newcommand{\be}{\begin{equation}\nonumber}
\newcommand{\ee}{\end{equation}}
\newcommand{\beq}{\begin{eqnarray}}
\newcommand{\eeq}{\end{eqnarray}}
\newcommand{\bem}{\begin{multline}}
\newcommand{\eem}{\end{multline}}

\renewcommand{\figurename}{Figure}
\renewcommand{\tablename}{Table}

\begin{document}

\title[Article Title]{Graded phononic metamaterials: Scalable design meets scalable microfabrication}

\author[1,2]{\fnm{Charles} \sur{Dorn}}\email{cdorn@uw.edu}
\equalcont{These authors contributed equally to this work.}

\author[1,3]{\fnm{Vignesh} \sur{Kannan}}\email{vignesh.kannan@polytechnique.edu}
\equalcont{These authors contributed equally to this work.}

\author[4]{\fnm{Ute} \sur{Drechsler}}\email{dre@zurich.ibm.com}

\author*[1]{\fnm{Dennis M.} \sur{Kochmann}}\email{dmk@ethz.ch}

\affil*[1]{\orgdiv{Mechanics and Materials Laboratory}, \orgname{ETH Zurich}, \orgaddress{\city{Zurich}, \postcode{8092}, \country{Switzerland}}}

\affil[2]{\orgdiv{Department of Aeronautics and Astronautics}, \orgname{University of Washington}, \orgaddress{\city{Seattle}, \postcode{98195}, \state{WA}, \country{USA}}}

\affil[3]{\orgdiv{Laboratoire de Mécanique des Solides}, \orgname{École Polytechnique}, \orgaddress{\city{Palaiseau}, \postcode{91128}, \country{France}}}

\affil[4]{\orgname{IBM Research -- Zurich}, \orgaddress{\city{Ruschlikon}, \postcode{8803}, \country{Switzerland}}}

\abstract{Metamaterials are a new generation of advanced materials, exhibiting engineered microstructures that enable customized material properties not found in nature. The dynamics of metamaterials are particularly fascinating, promising the capability to guide, attenuate, and focus  waves at will. Phononic metamaterials aim to manipulate mechanical waves with broad applications in acoustics, elastodynamics, and structural vibrations. A key bottleneck in the advancement of phononic metamaterials is scalability -- in design, simulation, and especially fabrication (e.g., beyond tens of unit cells per spatial dimension). We present a framework for scalable inverse design of spatially graded metamaterials for elastic wave guiding, together with a scalable microfabrication method. This framework enables the design and realization of complex waveguides including hundreds of thousands of unit cells, with the potential to extend to millions with no change in protocol. Scalable design is achieved via optimization with a ray tracing model for waves in spatially graded beam lattices. Designs are fabricated by photolithography and etching of silicon wafers to create free-standing microarchitected films. Wave guiding is demonstrated experimentally, using pulsed laser excitation and an interferometer for displacement measurements. Broadband wave guiding is demonstrated, indicating the promise of our scalable design and fabrication methods for on-chip elastic wave manipulation.
}

\maketitle

\section*{Introduction}

The nature of wave propagation in periodic lattices drives the fundamental behavior of materials, from the electronic band structure to thermal and optical properties. Phononic metamaterials build a bridge from wave phenomena at the atomic \cite{brillouin1946wave} to engineering scales, where lattices can be carefully designed to manipulate mechanical waves \cite{hussein2014dynamics}. As rapidly advancing manufacturing technology has enabled the fabrication of extremely intricate geometries, a rich literature on metamaterials has emerged in pursuit of harnessing mechanical waves for broad applications including vibration and acoustic isolation \cite{brule2014experiments,manushyna2023application}, sensing \cite{dubvcek2024sensor}, and energy harvesting \cite{chen2014metamaterials}.

A key limitation that plagues the advancement of metamaterials is the scalability of both computation and fabrication. It is challenging to model, design, and manufacture large architectures spanning thousands to millions of unit cells, which truly dissolve the differences between materials and structures. Solving the problem of scalability promises a significantly enlarged and widely untapped design space for programmable control of (meta-)material functionality to enable emerging engineering applications.

On the computational side, the efficiency of modeling and design methods is a bottleneck to scalability due to the multiscale nature of metamaterials. While efficient multiscale modeling and design methods are available for periodic architectures thanks to Bloch's theorem \cite{brillouin1946wave}, periodic architectures barely scratch the surface of the vast metamaterial design space. Looking beyond periodic architectures (e.g., by a spatial grading that smoothly varies the unit cell architecture in space), promises enhanced functionalities such as wave focusing \cite{tol2017phononic}, broadband attenuation \cite{zhu2013acoustic}, and signal processing \cite{dorn2023conformally}. However, modeling and design of non-periodic architectures are difficult to scale to large architectures with many unit cells since, unlike for periodic architectures, each unit cell must be modeled. Existing spatial grading design methods have relied on heavily restricted design spaces (e.g., linear gradings \cite{trainiti2016wave} and analytical solutions \cite{lin2009gradient}) or restricting assumptions of long wavelengths to enable homogenization \cite{chen2010acoustic,nassar2020polar}. Thus, there is an immensely rich but largely unexplored design space of spatially variant metamaterials spanning large numbers of unit cells.

On the experimental side, the fabrication of metamaterials consisting of large numbers of unit cells is challenging. Scaling fabrication beyond tens of unit cells per dimension is extremely challenging, making it difficult to truly fabricate meta-\textit{materials} rather than structures. At the current limits of 3D printing, architectures with hundreds of unit cells per dimension \cite{shaikeea2022toughness} are within reach, but such methods are limited to polymeric materials, which are not ideal for wave guiding due to material damping. Scalability is especially challenging for nano- and microscale manufacturing, where two-photon lithography \cite{lee2012micro,buckmann2014elasto,bauer2017nanolattices,meza2017reexamining,harinarayana2021two,krodel2016microlattice} and its variants \cite{kiefer2024multi} can hardly scale beyond samples with tens of unit cells per dimension in a tractable build time. Microfabrication of optical metamaterials has achieved better scalability, with two-dimensional architectures reaching over a billion unit cells, using electron beam lithography and atomic layer deposition \cite{li2022inverse,li2022empowering}; however, this is achieved by patterning features onto a substrate, rather than creating a free-standing material architecture.

In this work, we introduce a solution to metamaterial scalability both computationally and experimentally. Our results enable the design and realization of elaborate spatially graded metamaterials spanning at least three orders of magnitude in length scales with tens of thousands of unit cells. To achieve scalable computational design, we leverage an optimization framework based on ray tracing for efficient modeling of wave propagation. A modular design approach is developed, where multiple \textit{tiles} are independently designed to achieve different wave guiding functionalities. The tiles act as building blocks that are assembled like puzzle pieces into large designs with customizable wave guiding capabilities. To realize our designs, microfabrication based on photolithography and etching of silicon wafers creates free-standing architected films. Experimental demonstration of wave guiding is achieved using a nanosecond pulsed laser for excitation and heterodyne interferometry for measurement of particle displacements. The demonstrated design and microfabrication concept opens doors to new applications such as on-chip vibration isolation and signal processing for microscale electromechanical systems (MEMS).

\begin{figure*}[t!]
    \centering
    \includegraphics[width=\linewidth]{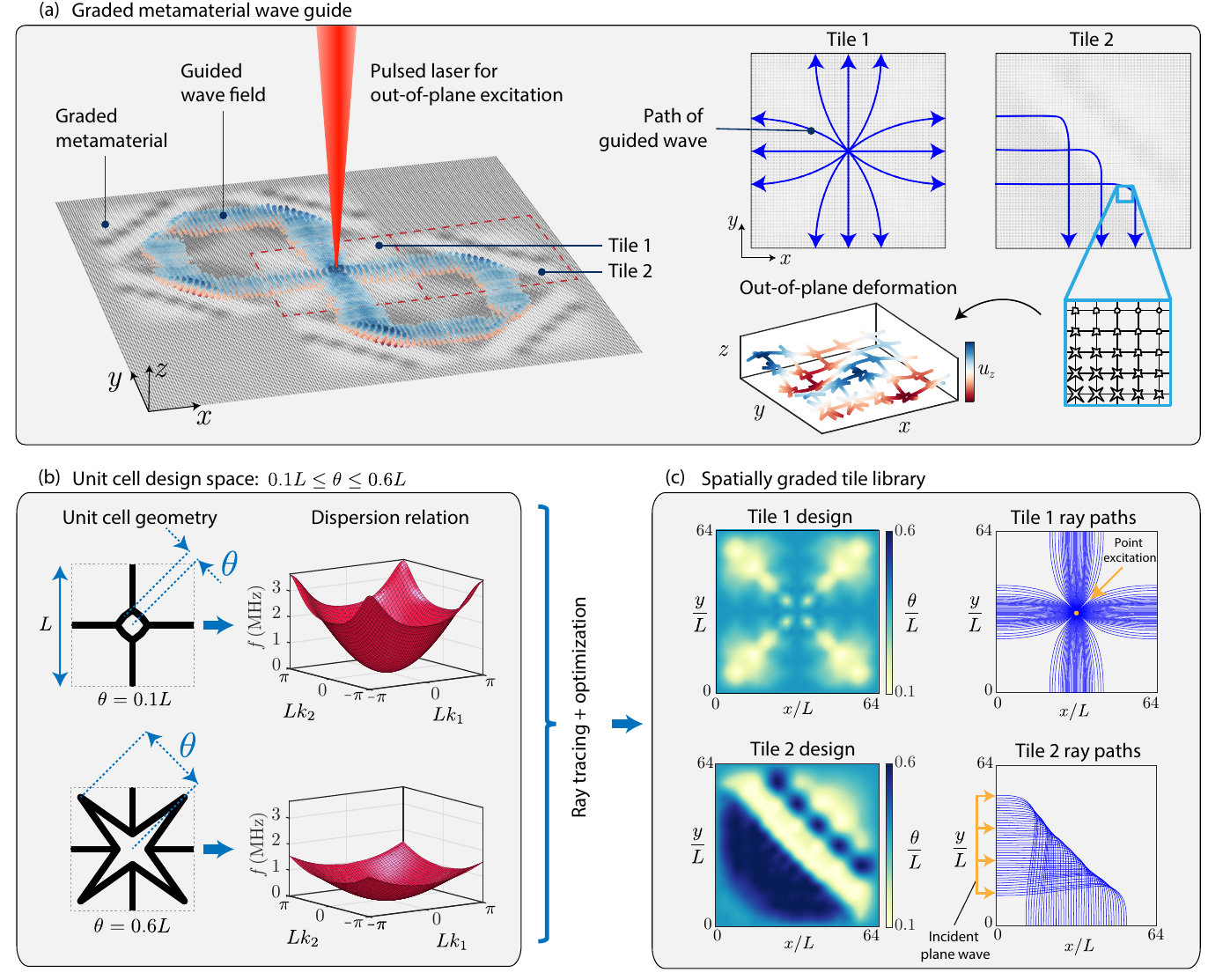}
    \caption{\textbf{Spatially-graded metamaterial design concept}: (a) Schematic illustration of the metamaterial composed of tiles, whose wave properties are carefully designed through the spatially varying unit cell geometry. (b) The unit cell geometry is modulated by varying $\theta$ (shown are two example configurations and the corresponding wave dispersion relations). (c) Library of tile designs and the corresponding wave motion illustrated by ray trajectories: Tile~1 splits the wave emanating from a point excitation, while Tile~2 redirects a plane wave by $90^\circ$.}
    \label{fig:figure1}
\end{figure*}

\section*{Modular waveguide design}

Spatially grading the properties of a material is a powerful tool for manipulating propagating waves, which has long been used, e.g., for gradient-index optics \cite{moore1980gradient}. Extending this idea to grading the unit cells of phononic metamaterials introduces a multiscale problem, which is computationally challenging to simulate and optimize. Brute-force transient simulations (e.g., using finite elements) are accurate but prohibitively slow for inverse design of large microstructures. Hence, existing work has relied primarily on simple, intuitive designs (such as radially symmetric and linear gradings \cite{tol2017phononic,trainiti2016wave,jin2019gradient}), leaving a rich untapped design space of complex spatial gradings to be explored. 

To circumvent the computational efficiency bottleneck, we leverage ray tracing to model wave motion in graded metamaterials, generalizing well-established ray theories for smooth continua (e.g., seismic ray theory \cite{cerveny2001seismic} and geometric optics \cite{born2013principles}) to graded metamaterials \cite{dorn2022ray}. Ray tracing in metamaterials provides an efficient modeling tool, which relies on \textit{local} Bloch wave analysis in the neighborhood of each unit cell to compute \textit{local} dispersion relations (assuming smooth spatial gradings and hence locally an approximately periodic medium). Consequently, this method applies both within and above the low-frequency homogenization limit, since the complete dispersion relations are accounted for. The resulting spatially variant local dispersion relations act as a Hamiltonian for tracing rays to determine how waves propagate in graded metamaterials.

\begin{figure*}[t]
    \centering
    \includegraphics[width=\linewidth]{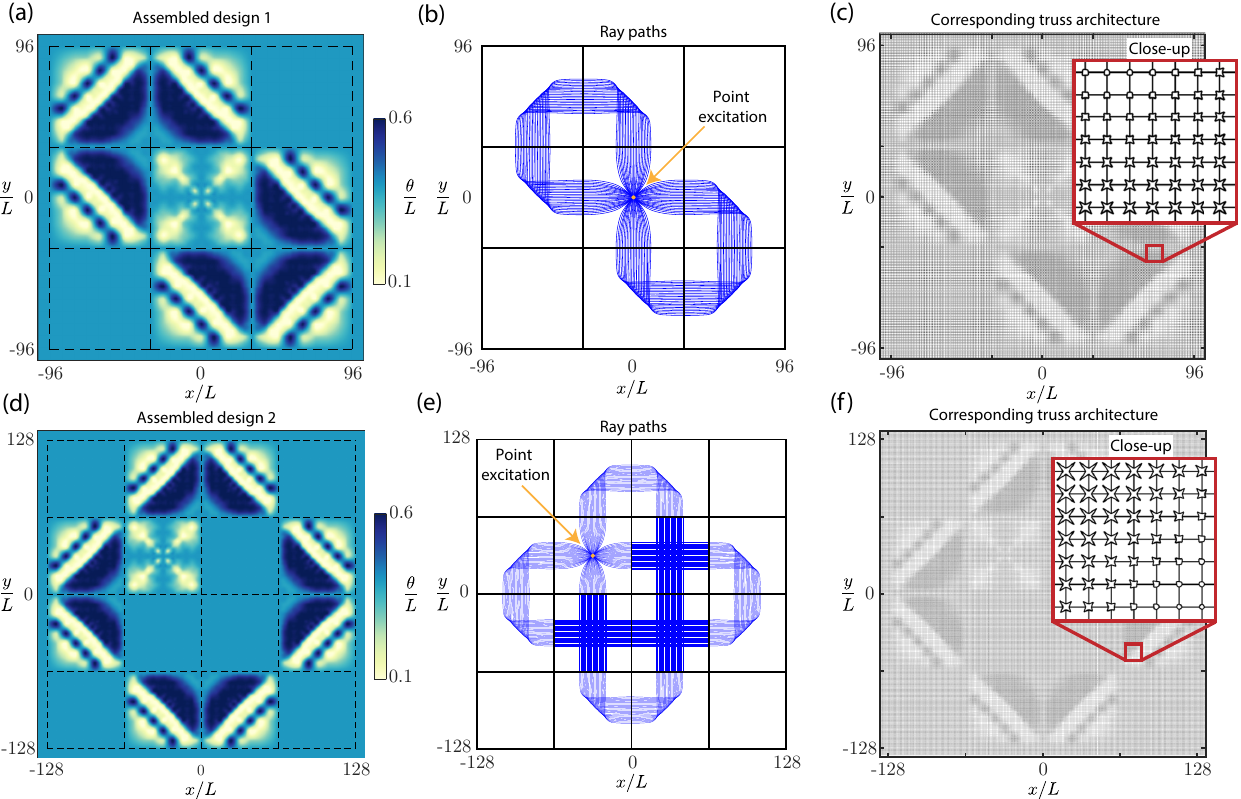}
    \caption{\textbf{Modular design by tile assembly:} (a) Spatial distribution of design parameter~$\theta$ for an assembly of $3\times 3$ tiles. (b) Ray trajectories for the assembly in (a) in response to a point excitation at the center. (c) Graded metamaterial containing $192\times192$ unit cells, realizing the design distribution in (a) to result in the ''figure-eight'' wave motion in (b). (d) Spatial distribution of design parameter~$\theta$ for an assembly of $4\times 4$ tiles. (e) Ray trajectories for the assembly in (d) in response to a point excitation. (f) Graded metamaterial containing $256\times256$ unit cells, realizing the design distribution in (d) to result in the complex wave motion in (e).}
    \label{fig:figure2}
\end{figure*}

In this work, we adopt ray tracing for efficient forward modeling, which is crucial to achieve scalability to large and complicated graded geometries. An optimization-based inverse design framework is built around ray tracing by extending our previous formulations~\cite{dorn2023inverse}. Specifically, an optimization problem is posed to design the spatial distribution of unit cells such that the corresponding ray paths are shaped in a prescribed way. Consider the unit cells in Figure~\ref{fig:figure1}b, in which each edge represents an elastic beam. The unit cell has one design parameter~$\theta$, which is restricted to the range $0.1L\leq \theta \leq 0.6L$ for manufacturability ($L$ is the height and width of the square unit cell). The foundation for forward modeling using ray tracing is the local dispersion relation throughout the metamaterial, which is computed using a beam finite element model \cite{zelhofer2017acoustic,telgen2024rainbow}. Focusing on out-of-plane bending vibrations as in our experiments, the lowest dispersion surface is plotted in Figure~\ref{fig:figure1}b for the two extremes of the design space. For this unit cell, the lowest out-of-plane dispersion surface does not intersect other out-of-plane dispersion surfaces, so it can be studied in isolation. We aim to optimally design the spatial distribution of $\theta$ in a metamaterial consisting of many unit cells to achieve a prescribed wave guiding objective through the resulting spatial variation of the (local) dispersion relations. 

An optimization problem is solved to shape the ray trajectories in two spatially graded square tiles, shown in Figure~\ref{fig:figure1}c, each spanning 64$\times$64 unit cells. The first design, Tile~1, considers a point excitation at the center unit cell (from which the rays originate) and designs the spatial distribution of $\theta$ to guide the resulting wave to exit the tile along the four tile edges with rays perpendicular to the edges. The second tile, Tile~2, considers an incident plane wave, with horizontal rays entering from the left tile boundary, and redirects the rays to exit perpendicular to the bottom tile edge, thus rotating the incident plane wave by $90^\circ$. Both tiles are designed to have $\theta=0.4L$ around the perimeter, to ensure compatibility between tiles when assembled into the metamaterial (Figure~\ref{fig:figure1}a). The distribution of $\theta$ and the corresponding ray trajectories are plotted in Figure \ref{fig:figure1}c for both tiles. Details of the optimization setup for tile design are presented in the Supplementary Information Section 1.

Since the two tiles are designed to be compatible, there is no jump in $\theta$ at the shared boundary when the two tiles are placed next to each other. Consequently, ray paths of Figure~\ref{fig:figure1}c can be continuously connected between adjacent tiles. This allows for solutions to be ``assembled" by placing tiles next to each other such that the rays of each tile are connected to form a desired set of ray paths. Figure~\ref{fig:figure2}a provides an example of a tile assembly, which plots the spatial distribution of $\theta$ with dotted lines highlighting the boundary between tiles. The corresponding ray trajectories are shown in Figure~\ref{fig:figure2}b, which are connected to form a figure-eight shape. Thus, waves emerging from out-of-plane excitation at the origin are guided along the figure-eight. The geometric realization of this assembled design is shown in Figure~\ref{fig:figure2}c, where the geometry of each unit cell is determined from the designed distribution of $\theta$ in Figure~\ref{fig:figure2}a. Note that, since all rays in the figure-eight return to the point of excitation, the design -- in the ideal case without dissipative losses -- would result in continued traversal of the figure-eight even after the point excitation is removed.

Design by tile assembly is a scalable approach that leads to many possible designs. A second example of assembled tiles is shown in Figure~\ref{fig:figure2}d. Its ray trajectories guide the wave emerging from a point excitation along the outline of a cross, as shown in Figure~\ref{fig:figure2}e. The corresponding beam architecture is shown in Figure~\ref{fig:figure2}f. Transient finite element simulations are performed (see Supplementary Information Section 2) to validate the designs. Supplementary Videos 1 and 2 show animations of the finite element simulation results for the case of harmonic loading at the design frequency. We note that ray trajectory design is performed for a specific target frequency but -- in both simulations and experiments -- wave guiding is observed over a broad frequency band surrounding this target frequency (see the Supplementary Information Sections 2 and 5 for details).

This modular approach offers a means of designing broadband metamaterial waveguides spanning a large number of unit cells, taking full advantage of spatial grading. The examples of Figure~\ref{fig:figure2}a and \ref{fig:figure2}d span approximately $37{,}000$ and $66{,}000$ unit cells, respectively, and this approach directly scales to larger designs simply by assembling more tiles. Thus, through efficient modeling and optimization based on ray tracing together with a modular tile assembly approach, we can circumvent the challenges of computational scalability to compute large and elaborate metamaterial designs. The challenge that remains is scalable fabrication to realize such designs, for which we turn to semiconductor microfabrication, drawing inspiration from chip manufacturing methods.

\section*{Microfabrication: Silicon wafers to beam-based metamaterials}

Fabrication of spatially graded architectures spanning multiple length scales, such as those in Figure~\ref{fig:figure2}, is out of reach of standard manufacturing methods at the macroscale. We adopt silicon microfabrication techniques for a solution. 

\begin{figure*}[h]
    \centering
    \includegraphics[width=\linewidth]{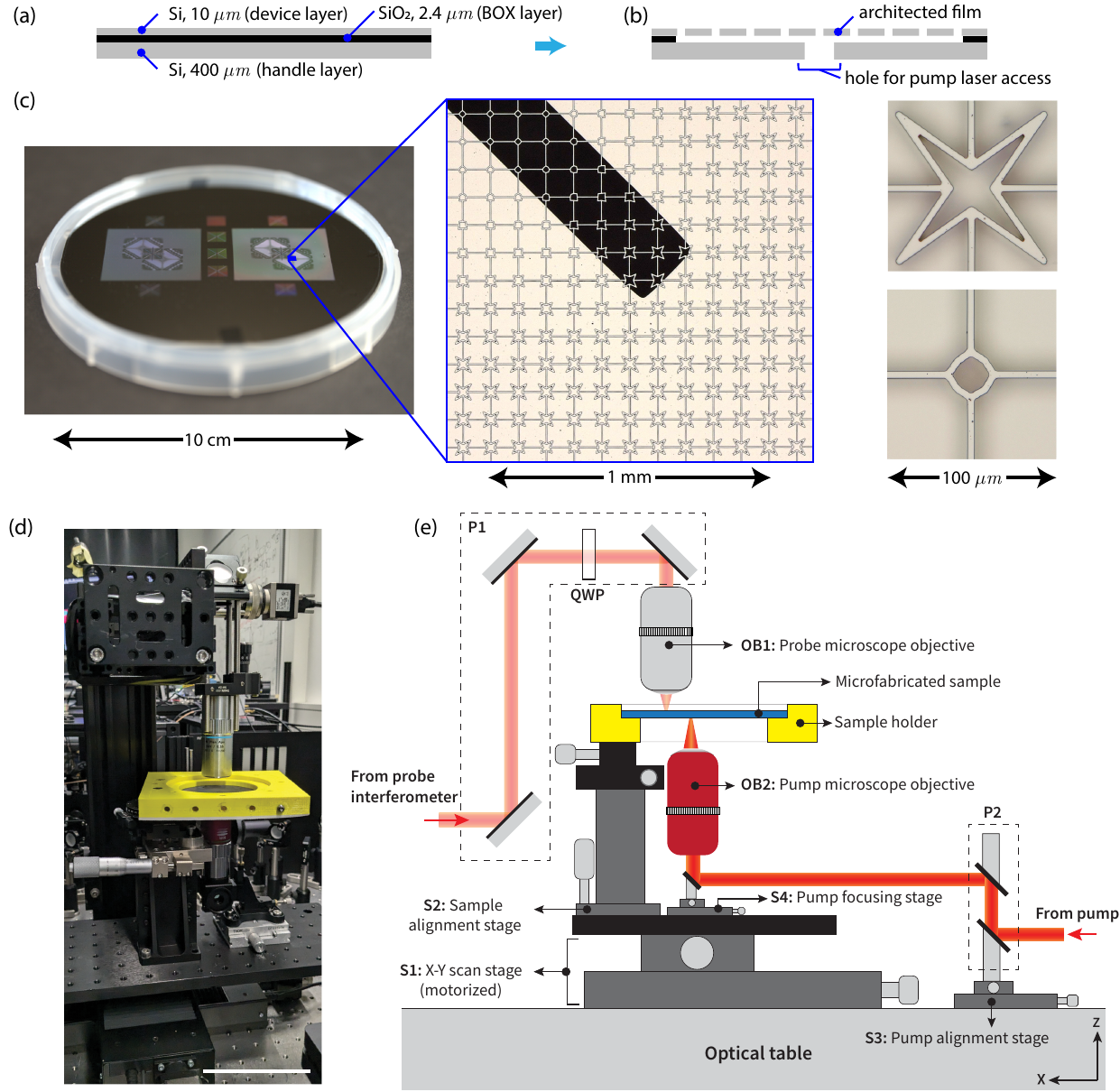}
    \caption{\textbf{Fabrication and experimental setup:} Schematic of the SOI wafer sample (a) before and (b) after microfabrication (layer thickness is not drawn to scale). (c) Micrographs of a prototype, showing the full wafer, a magnified top view of $13\times13$ unit cells (with a bottom hole visible in black), and two example unit cells corresponding to the two design space extremes of Figure~\ref{fig:figure2}a. (d) Photograph and (e) schematic of the experimental setup used to probe wave motion in wafer-based graded metamaterials.}
    \label{fig:figure3}
\end{figure*}

The graded structure of Figure~\ref{fig:figure2}c is fabricated in the top (device) layer of a silicon-on-insulator (SOI) wafer. A schematic of the cross-section of an as-received SOI wafer is shown in Figure~\ref{fig:figure3}a. Our fabrication process involves photolithography and deep reactive ion etching to architect the device layer, followed by removal of the buried oxide (SiO$_2$) layer, using vapor hydrofluoric acid etching. This results in a free-standing architected film supported by the substrate only around the outer perimeter, similar to a drum with an architected membrane. Additionally, windows are etched in the supporting substrate to enable remote excitation of the acoustic wave from the bottom with a wide range of potential spatio-temporal profiles. To enable optimal excitation and measurement of wave propagation in these films, thin aluminum transducers ($\sim 30$~nm thick) were vapor-deposited on both sides of the architected film. 

A schematic of the wafer cross-section after fabrication is shown in Figure~\ref{fig:figure3}b and micrographs of the prototype in Figure~\ref{fig:figure3}c. The novelty of this microfabrication method is that free-standing architected films (as opposed to a pattern deposited on a substrate) are manufacturable in a scalable fashion. That is, the entire wafer can be architected with micron-scale features, with the only scalability limitation being the size of the wafer itself. We have demonstrated this scalability by manufacturing $\sim 600,000$ unit cells in a single SOI wafer of 100~mm diameter~\cite{kannan2025}. 

\section*{Experimental wave guiding demonstration}

Experiments were performed to demonstrate wave guiding in the prototype, using a pump-probe setup. A 1030~nm nanosecond pulsed laser was used for photoacoustic pump excitation, sending broadband out-of-plane elastic wave modes through the sample. The out-of-plane displacement response was measured by a custom-built heterodyne interferometer. A photograph and schematic of the experimental setup are shown in Figure~\ref{fig:figure3}d-e. Due to the repeatability of elastic wave propagation, data is collected by successively pumping the excitation pulse at the same spatial location, while the probe laser scans different positions on the wafer. The signal-to-noise ratio was excellent at frequencies up to 2~MHz and measurable up to 4~MHz. This was possible due to low intrinsic damping of single-crystal silicon and the high displacement resolution of the interferometer. Though not necessary here, any dissipation arising from the surrounding air can be mitigated by experiments in vacuum. The experiment hence has a unique potential to probe wave attenuation due to architecture alone. 

\begin{figure*}[!ht]
    \centering
    \includegraphics[width=\linewidth]{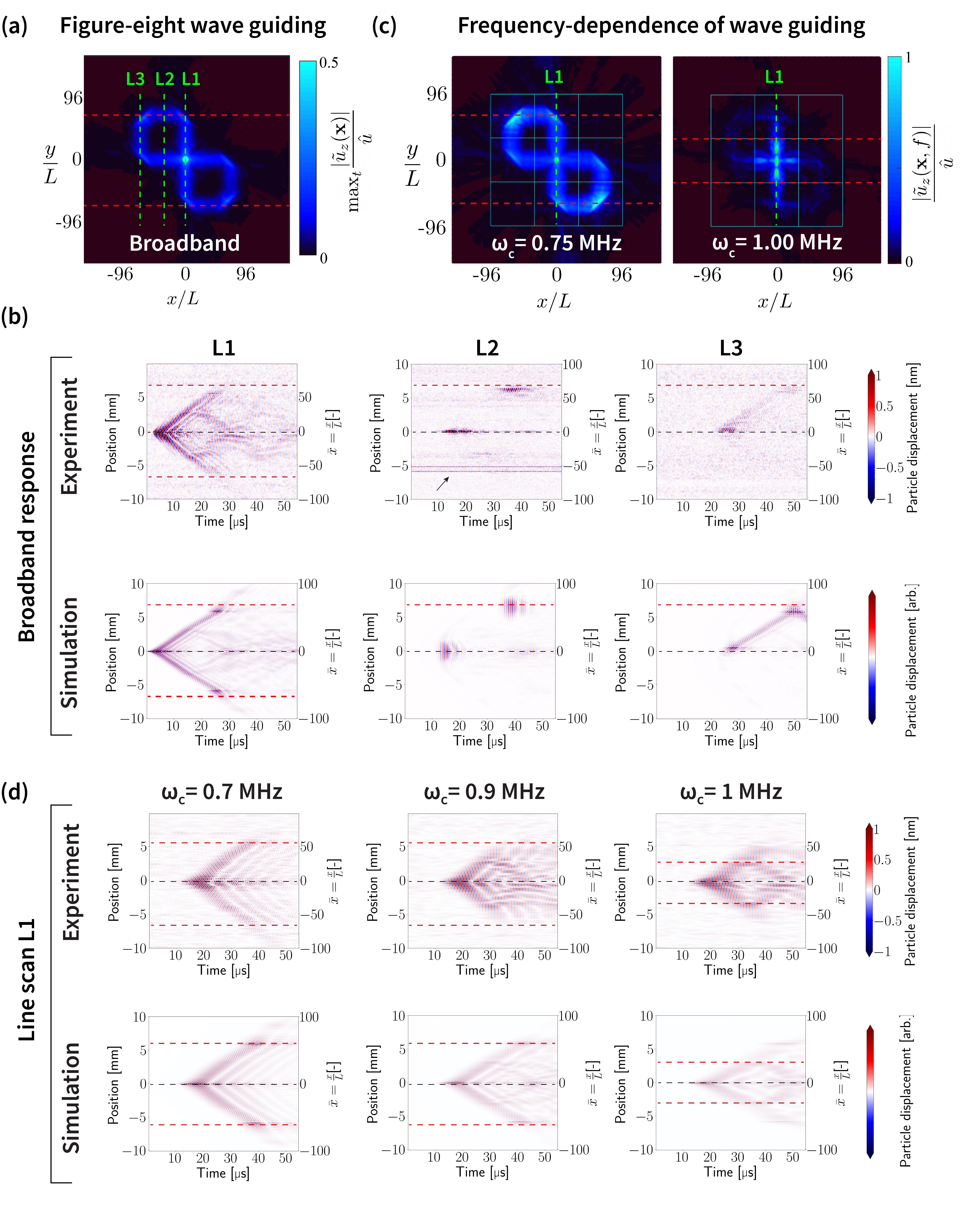}
    \caption{\textbf{Experimental results showing figure-eight wave guidance}. Finite element simulation results for (a)~transient wave propagation and (c)~frequency-dependent wave guiding in the graded architecture. Green dashed lines indicate the positions of line scans collected during experiments. (b)~Experimental results from scans L1, L2, and L3, in comparison with equivalent data from finite element simulations, showing an excellent match. Red dashed lines indicate positions beyond which no wave propagation was measured/calculated (equivalent positions are shown in (a)). (d)~Comparisons of frequency-dependent responses between experiments and simulations for L1 confirm agreement across a wide frequency range. 
    }
    \label{fig:figure4}
\end{figure*}

To capture the metamaterial sample's ability to guide elastic waves, data is collected along three lines, at $x=0$ (denoted L1), $x=-32L$ (L2), and $x=-64L$ (L3), see Figure~\ref{fig:figure4}a. At 100 measurement points along each line, the time series displacement signal is measured immediately following a pump excitation. Pump excitation occurs at the same point on the sample for all measurement scans. Figure~\ref{fig:figure4} shows the measured response at each position along L1, L2, and L3, comparing the experimental results to simulated data from a finite element model (see Supplementary Information Section 2). The signal-to-noise ratio was excellent for all measurement points except two (marked by the black arrow for line scan L2). For reference, Figure~\ref{fig:figure4}a shows the maximum displacement at each spatial location during a transient finite element simulation, indicating each of the line scans with respect to the guided wave path.

The measured response of the wafer clearly captures wave guiding along the designed figure-eight trajectory. The experimental data closely matches the finite element simulation data, with agreement in both the location and timing of large displacement amplitudes. Furthermore, while the waveguide is designed for a specific target frequency of 750~kHz, broadband wave guiding is observed. A detailed frequency analysis is presented in Supplementary Information Section 5, showing that wave guiding is achieved in the window of 250-800 kHz, both in experimental and simulated data. The experiment further shows a different wave guiding response above 800~kHz, which has been validated using finite element analysis. The latter was not explicitly introduced during inverse design. Thus, the proposed methodology is capable of customized manipulation and discovery of phenomena across a frequency range broader than that of the inverse design. The observed broadband wave guiding likely stems from having dispersion relations with approximately self-similar isofrequency contours and angular distributions of group velocity (which drive ray trajectories) over the frequency range 250-800~kHz for all unit cells in the chosen design space.

\section*{Conclusion}

This work pushes the limits of metamaterial scalability in computational design and experiments. We have presented an inverse design framework together with a microscale wafer fabrication method, both of which are demonstrated on designs spanning hundreds of unit cells per dimension with potential for further scalability. Our design method relies on optimization of an efficient ray tracing model of wave motion in spatially graded metamaterials. A versatile modular approach involves designing individual tiles that are assembled to achieve complex wave guiding objectives. This leads to designs spanning multiple length scales. To realize those designs, we adapt wafer manufacturing methods to create free-standing architected films, with the scalability to fill an entire wafer with millions of unit cells. Experiments based on a pump-probe scheme for excitation and measurement demonstrate broadband wave guiding capabilities of an optimized design.

Unlocking scalability of acoustic metamaterials promises both scientific and technological advances. On the scientific side, the presented methods enable a true separation of length scales to access the \textit{material} rather than \textit{structural} regime, providing a valuable setting for future experimentation. Furthermore, the proposed microfabrication technique paves the way for high-throughput experimentation by fabrication of many samples on a single wafer. On the technological side, the presented scalable approaches greatly expand the design space of metamaterials, consequently expanding their functionality for applications from vibration isolation in MEMS \cite{zega2022microstructured} to high-frequency energy harvesting \cite{hu2021acoustic} to microfluidics \cite{charara2025spatially}, also leveraging the demonstrated broadband stability of the waveguide. The presented waveguide designs serve as illustrative examples, in which wave guiding along customized trajectories is enabled by scalability to large designs. 

\backmatter

\bmhead{Methods}
\subsection*{Microfabrication}

We use cleanroom-based microfabrication processes on commercially-procured 100~mm Silicon-On-Insulator (SOI) wafers to manufacture the figure-eight design prototype of Figure~2a-c. The procedure involves a combination of optical lithography and deep reactive ion etching (DRIE) on the top and bottom layers of the SOI wafers, followed by vapor hydrofluoric acid (HF) etching to remove the intermediate layer. The final results are free-standing architected films supported by a 400~\textmu m thick base substrate layer. A detailed description of the microfabrication procedure is presented in the Supplementary Information Section 3.

Two samples are fabricated on a single wafer to reduce the number of wafers used. The figure-eight design prototype spans a 3$\times$3 cm region with a square unit cell of $L=100 \, \mu$m side length. Individual beams are $5 \, \mu$m wide. To enable excitation via pulsed laser in the experiments, holes are etched in the substrate layer to allow the pulsed laser to reach the device layer from the backside. 

\section*{Experimental characterization}

A photoacoustic pump-probe experiment was developed to resolve propagating elastic waves in our architected films. Excitation of the acoustic waves was achieved by an infrared pulsed laser beam (Coherent FLARE-NX, wavelength 1030~nm, pulse energy 500~\textmu J, pulse width~1~ns). The mechanism of photoacoustic excitation involves rapid thermal expansion of thin aluminum films deposited on the sample (acting as transducers), resulting in the propagation of an acoustic pulse through the film. The duration of this pulse is determined, in part, by the thickness of the film and its bulk elastic wave speed. 
Particle displacements were measured by a custom-built heterodyne interferometer. The heterodyne system uses a continuous wave laser (20~mW 633~nm) source with the reference branch frequency shifted by 80~MHz, using an Acousto-Optic Modulator (AOM: EQ Photonics 3080-120) driven by an RF driver (Gooch and Housego 3910). The sample branch of the interferometer was focused on the measurement point on the sample, using a 20X objective lens before being recombined and detected at a balanced photo detector (ThorLabs PDB 230A). Under ambient conditions, this generates an 80~MHz beat signal, which undergoes phase shifts due to displacement of the measurement point. During the propagation of the acoustic wave, the time-resolved phase shift was measured using a lock-in amplifier (Zurich Instruments GHFLI) and read out into a high-speed digital oscilloscope (Tektronix MSO64B). Particle displacements are directly proportional to the measured phase shift ($\varphi_{m}$), and are calculated as
\begin{equation}
    \delta(t) = \dfrac{\lambda}{2 \pi} \varphi_{m}(t),
\end{equation}
where $\lambda$ is the wavelength of the laser source. Scanning measurements were performed using two automated stages remotely controlled using Python code. At each measurement point, 50 time series data were averaged to increase the signal-to-noise ratio. Further details regarding data analysis are summarized in Supplementary Information Sections 4 and 5.

\bmhead{Acknowledgements}
The authors thank Dr.~Emil Bronstein for his assistance in setting up experimental scans. C.D.\ acknowledges partial support from an ETH Zurich Postdoctoral Fellowship. Mask writing and fabrication were performed in the cleanroom facility of the Binnig and Rohrer Nanotechnology Center of IBM Zurich.

\bmhead{Author contribution}
C.D., V.K., and D.M.K.\ designed the research. C.D.\ performed computational modeling and design. V.K.\ designed the microfabrication method and experiments. V.K.\ and C.D.\ performed the experiments and analysis. U.D.\ supported the microfabrication efforts. C.D., V.K., and D.M.K.\ wrote the manuscript.

\bmhead{Data availability}
Data and codes supporting this study are publicly available in a repository accessible at \url{https://doi.org/10.3929/ethz-b-000742467}.

\bibliography{bibliography}

\begin{thebibliography}{10}
\expandafter\ifx\csname url\endcsname\relax
  \def\url#1{\burl{#1}}\fi
\expandafter\ifx\csname urlprefix\endcsname\relax\def\urlprefix{URL }\fi
\providecommand{\bibinfo}[2]{#2}
\providecommand{\eprint}[2][]{\url{#2}}
\providecommand{\doi}[1]{\url{https://doi.org/#1}}
\bibcommenthead

\bibitem{brillouin1946wave}
\bibinfo{author}{Brillouin, L.}
\newblock \bibinfo{title}{Wave propagation in periodic structures}.
\newblock \emph{\bibinfo{journal}{McGraw-Hill}} \textbf{\bibinfo{volume}{2}}
  (\bibinfo{year}{1946}).

\bibitem{hussein2014dynamics}
\bibinfo{author}{Hussein, M.~I.}, \bibinfo{author}{Leamy, M.~J.} \&
  \bibinfo{author}{Ruzzene, M.}
\newblock \bibinfo{title}{Dynamics of phononic materials and structures:
  Historical origins, recent progress, and future outlook}.
\newblock \emph{\bibinfo{journal}{Applied Mechanics Reviews}}
  \textbf{\bibinfo{volume}{66}}, \bibinfo{pages}{040802}
  (\bibinfo{year}{2014}).

\bibitem{brule2014experiments}
\bibinfo{author}{Br{\^u}l{\'e}, S.}, \bibinfo{author}{Javelaud, E.},
  \bibinfo{author}{Enoch, S.} \& \bibinfo{author}{Guenneau, S.}
\newblock \bibinfo{title}{Experiments on seismic metamaterials: molding surface
  waves}.
\newblock \emph{\bibinfo{journal}{Physical Review Letters}}
  \textbf{\bibinfo{volume}{112}}, \bibinfo{pages}{133901}
  (\bibinfo{year}{2014}).

\bibitem{manushyna2023application}
\bibinfo{author}{Manushyna, D.} \emph{et~al.}
\newblock \bibinfo{title}{Application of vibroacoustic metamaterials for
  structural vibration reduction in space structures}.
\newblock \emph{\bibinfo{journal}{Mechanics Research Communications}}
  \textbf{\bibinfo{volume}{129}}, \bibinfo{pages}{104090}
  (\bibinfo{year}{2023}).

\bibitem{dubvcek2024sensor}
\bibinfo{author}{Dub{\v{c}}ek, T.} \emph{et~al.}
\newblock \bibinfo{title}{In-sensor passive speech classification with phononic
  metamaterials}.
\newblock \emph{\bibinfo{journal}{Advanced Functional Materials}}
  \textbf{\bibinfo{volume}{34}}, \bibinfo{pages}{2311877}
  (\bibinfo{year}{2024}).

\bibitem{chen2014metamaterials}
\bibinfo{author}{Chen, Z.}, \bibinfo{author}{Guo, B.}, \bibinfo{author}{Yang,
  Y.} \& \bibinfo{author}{Cheng, C.}
\newblock \bibinfo{title}{Metamaterials-based enhanced energy harvesting: A
  review}.
\newblock \emph{\bibinfo{journal}{Physica B: Condensed Matter}}
  \textbf{\bibinfo{volume}{438}}, \bibinfo{pages}{1--8} (\bibinfo{year}{2014}).

\bibitem{tol2017phononic}
\bibinfo{author}{Tol, S.}, \bibinfo{author}{Degertekin, F.~L.} \&
  \bibinfo{author}{Erturk, A.}
\newblock \bibinfo{title}{Phononic crystal luneburg lens for omnidirectional
  elastic wave focusing and energy harvesting}.
\newblock \emph{\bibinfo{journal}{Applied Physics Letters}}
  \textbf{\bibinfo{volume}{111}} (\bibinfo{year}{2017}).

\bibitem{zhu2013acoustic}
\bibinfo{author}{Zhu, J.} \emph{et~al.}
\newblock \bibinfo{title}{Acoustic rainbow trapping}.
\newblock \emph{\bibinfo{journal}{Scientific Reports}}
  \textbf{\bibinfo{volume}{3}}, \bibinfo{pages}{1728} (\bibinfo{year}{2013}).

\bibitem{dorn2023conformally}
\bibinfo{author}{Dorn, C.} \& \bibinfo{author}{Kochmann, D.~M.}
\newblock \bibinfo{title}{Conformally graded metamaterials for elastic wave
  guidance}.
\newblock \emph{\bibinfo{journal}{Extreme Mechanics Letters}}
  \textbf{\bibinfo{volume}{65}}, \bibinfo{pages}{102091}
  (\bibinfo{year}{2023}).

\bibitem{trainiti2016wave}
\bibinfo{author}{Trainiti, G.}, \bibinfo{author}{Rimoli, J.~J.} \&
  \bibinfo{author}{Ruzzene, M.}
\newblock \bibinfo{title}{Wave propagation in undulated structural lattices}.
\newblock \emph{\bibinfo{journal}{International Journal of Solids and
  Structures}} \textbf{\bibinfo{volume}{97}}, \bibinfo{pages}{431--444}
  (\bibinfo{year}{2016}).

\bibitem{lin2009gradient}
\bibinfo{author}{Lin, S.-C.~S.}, \bibinfo{author}{Huang, T.~J.},
  \bibinfo{author}{Sun, J.-H.} \& \bibinfo{author}{Wu, T.-T.}
\newblock \bibinfo{title}{Gradient-index phononic crystals}.
\newblock \emph{\bibinfo{journal}{Physical Review B—Condensed Matter and
  Materials Physics}} \textbf{\bibinfo{volume}{79}}, \bibinfo{pages}{094302}
  (\bibinfo{year}{2009}).

\bibitem{chen2010acoustic}
\bibinfo{author}{Chen, H.} \& \bibinfo{author}{Chan, C.~T.}
\newblock \bibinfo{title}{Acoustic cloaking and transformation acoustics}.
\newblock \emph{\bibinfo{journal}{Journal of Physics D: Applied Physics}}
  \textbf{\bibinfo{volume}{43}}, \bibinfo{pages}{113001}
  (\bibinfo{year}{2010}).

\bibitem{nassar2020polar}
\bibinfo{author}{Nassar, H.}, \bibinfo{author}{Chen, Y.} \&
  \bibinfo{author}{Huang, G.}
\newblock \bibinfo{title}{Polar metamaterials: a new outlook on resonance for
  cloaking applications}.
\newblock \emph{\bibinfo{journal}{Physical Review Letters}}
  \textbf{\bibinfo{volume}{124}}, \bibinfo{pages}{084301}
  (\bibinfo{year}{2020}).

\bibitem{shaikeea2022toughness}
\bibinfo{author}{Shaikeea, A.}, \bibinfo{author}{Cui, H.},
  \bibinfo{author}{O’Masta, M.}, \bibinfo{author}{Zheng, X.~R.} \&
  \bibinfo{author}{Deshpande, V.~S.}
\newblock \bibinfo{title}{The toughness of mechanical metamaterials}.
\newblock \emph{\bibinfo{journal}{Nature Materials}}
  \textbf{\bibinfo{volume}{21}}, \bibinfo{pages}{297--304}
  (\bibinfo{year}{2022}).

\bibitem{lee2012micro}
\bibinfo{author}{Lee, J.-H.}, \bibinfo{author}{Singer, J.~P.} \&
  \bibinfo{author}{Thomas, E.~L.}
\newblock \bibinfo{title}{Micro-/nanostructured mechanical metamaterials}.
\newblock \emph{\bibinfo{journal}{Advanced Materials}}
  \textbf{\bibinfo{volume}{24}}, \bibinfo{pages}{4782--4810}
  (\bibinfo{year}{2012}).

\bibitem{buckmann2014elasto}
\bibinfo{author}{B{\"u}ckmann, T.}, \bibinfo{author}{Thiel, M.},
  \bibinfo{author}{Kadic, M.}, \bibinfo{author}{Schittny, R.} \&
  \bibinfo{author}{Wegener, M.}
\newblock \bibinfo{title}{An elasto-mechanical unfeelability cloak made of
  pentamode metamaterials}.
\newblock \emph{\bibinfo{journal}{Nature Communications}}
  \textbf{\bibinfo{volume}{5}}, \bibinfo{pages}{4130} (\bibinfo{year}{2014}).

\bibitem{bauer2017nanolattices}
\bibinfo{author}{Bauer, J.} \emph{et~al.}
\newblock \bibinfo{title}{Nanolattices: an emerging class of mechanical
  metamaterials}.
\newblock \emph{\bibinfo{journal}{Advanced Materials}}
  \textbf{\bibinfo{volume}{29}}, \bibinfo{pages}{1701850}
  (\bibinfo{year}{2017}).

\bibitem{meza2017reexamining}
\bibinfo{author}{Meza, L.~R.} \emph{et~al.}
\newblock \bibinfo{title}{Reexamining the mechanical property space of
  three-dimensional lattice architectures}.
\newblock \emph{\bibinfo{journal}{Acta Materialia}}
  \textbf{\bibinfo{volume}{140}}, \bibinfo{pages}{424--432}
  (\bibinfo{year}{2017}).

\bibitem{harinarayana2021two}
\bibinfo{author}{Harinarayana, V.} \& \bibinfo{author}{Shin, Y.}
\newblock \bibinfo{title}{Two-photon lithography for three-dimensional
  fabrication in micro/nanoscale regime: A comprehensive review}.
\newblock \emph{\bibinfo{journal}{Optics \& Laser Technology}}
  \textbf{\bibinfo{volume}{142}}, \bibinfo{pages}{107180}
  (\bibinfo{year}{2021}).

\bibitem{krodel2016microlattice}
\bibinfo{author}{Kr{\"o}del, S.} \& \bibinfo{author}{Daraio, C.}
\newblock \bibinfo{title}{Microlattice metamaterials for tailoring ultrasonic
  transmission with elastoacoustic hybridization}.
\newblock \emph{\bibinfo{journal}{Physical Review Applied}}
  \textbf{\bibinfo{volume}{6}}, \bibinfo{pages}{064005} (\bibinfo{year}{2016}).

\bibitem{kiefer2024multi}
\bibinfo{author}{Kiefer, P.} \emph{et~al.}
\newblock \bibinfo{title}{A multi-photon (7$\times$ 7)-focus 3d laser printer
  based on a 3d-printed diffractive optical element and a 3d-printed multi-lens
  array}.
\newblock \emph{\bibinfo{journal}{Light: Advanced Manufacturing}}
  \textbf{\bibinfo{volume}{4}}, \bibinfo{pages}{28--41} (\bibinfo{year}{2024}).

\bibitem{li2022inverse}
\bibinfo{author}{Li, Z.} \emph{et~al.}
\newblock \bibinfo{title}{Inverse design enables large-scale high-performance
  meta-optics reshaping virtual reality}.
\newblock \emph{\bibinfo{journal}{Nature Communications}}
  \textbf{\bibinfo{volume}{13}}, \bibinfo{pages}{1--11} (\bibinfo{year}{2022}).

\bibitem{li2022empowering}
\bibinfo{author}{Li, Z.}, \bibinfo{author}{Pestourie, R.},
  \bibinfo{author}{Lin, Z.}, \bibinfo{author}{Johnson, S.~G.} \&
  \bibinfo{author}{Capasso, F.}
\newblock \bibinfo{title}{Empowering metasurfaces with inverse design:
  principles and applications}.
\newblock \emph{\bibinfo{journal}{{ACS} Photonics}}
  \textbf{\bibinfo{volume}{9}}, \bibinfo{pages}{2178--2192}
  (\bibinfo{year}{2022}).

\bibitem{moore1980gradient}
\bibinfo{author}{Moore, D.~T.}
\newblock \bibinfo{title}{Gradient-index optics: a review}.
\newblock \emph{\bibinfo{journal}{Applied Optics}}
  \textbf{\bibinfo{volume}{19}}, \bibinfo{pages}{1035--1038}
  (\bibinfo{year}{1980}).

\bibitem{jin2019gradient}
\bibinfo{author}{Jin, Y.}, \bibinfo{author}{Djafari-Rouhani, B.} \&
  \bibinfo{author}{Torrent, D.}
\newblock \bibinfo{title}{Gradient index phononic crystals and metamaterials}.
\newblock \emph{\bibinfo{journal}{Nanophotonics}} \textbf{\bibinfo{volume}{8}},
  \bibinfo{pages}{685--701} (\bibinfo{year}{2019}).

\bibitem{cerveny2001seismic}
\bibinfo{author}{Cerven{\`y}, V.}
\newblock \emph{\bibinfo{title}{Seismic ray theory}}
  (\bibinfo{publisher}{Cambridge University Press Cambridge},
  \bibinfo{year}{2001}).

\bibitem{born2013principles}
\bibinfo{author}{Born, M.} \& \bibinfo{author}{Wolf, E.}
\newblock \emph{\bibinfo{title}{Principles of optics: {E}lectromagnetic theory
  of propagation, interference and diffraction of light}}
  (\bibinfo{publisher}{Elsevier}, \bibinfo{year}{2013}).

\bibitem{dorn2022ray}
\bibinfo{author}{Dorn, C.} \& \bibinfo{author}{Kochmann, D.~M.}
\newblock \bibinfo{title}{Ray theory for elastic wave propagation in graded
  metamaterials}.
\newblock \emph{\bibinfo{journal}{Journal of the Mechanics and Physics of
  Solids}} \textbf{\bibinfo{volume}{168}}, \bibinfo{pages}{105049}
  (\bibinfo{year}{2022}).

\bibitem{dorn2023inverse}
\bibinfo{author}{Dorn, C.} \& \bibinfo{author}{Kochmann, D.~M.}
\newblock \bibinfo{title}{Inverse design of graded phononic materials via ray
  tracing}.
\newblock \emph{\bibinfo{journal}{Journal of Applied Physics}}
  \textbf{\bibinfo{volume}{134}} (\bibinfo{year}{2023}).

\bibitem{zelhofer2017acoustic}
\bibinfo{author}{Zelhofer, A.~J.} \& \bibinfo{author}{Kochmann, D.~M.}
\newblock \bibinfo{title}{On acoustic wave beaming in two-dimensional
  structural lattices}.
\newblock \emph{\bibinfo{journal}{International Journal of Solids and
  Structures}} \textbf{\bibinfo{volume}{115}}, \bibinfo{pages}{248--269}
  (\bibinfo{year}{2017}).

\bibitem{telgen2024rainbow}
\bibinfo{author}{Telgen, B.} \emph{et~al.}
\newblock \bibinfo{title}{Rainbow trapping of out-of-plane mechanical waves in
  spatially variant beam lattices}.
\newblock \emph{\bibinfo{journal}{Journal of the Mechanics and Physics of
  Solids}} \textbf{\bibinfo{volume}{191}}, \bibinfo{pages}{105762}
  (\bibinfo{year}{2024}).

\bibitem{kannan2025}
\bibinfo{author}{Kannan, V.}, \bibinfo{author}{Dorn, C.},
  \bibinfo{author}{Drechsler, U.} \& \bibinfo{author}{Kochmann, D.}
\newblock \bibinfo{title}{Microscale architected materials for elastic wave
  guiding: Fabrication and dynamic characterization across length and time
  scales}  (\bibinfo{year}{2025}).

\bibitem{zega2022microstructured}
\bibinfo{author}{Zega, V.} \emph{et~al.}
\newblock \bibinfo{title}{Microstructured phononic crystal isolates from
  ultrasonic mechanical vibrations}.
\newblock \emph{\bibinfo{journal}{Applied Sciences}}
  \textbf{\bibinfo{volume}{12}}, \bibinfo{pages}{2499} (\bibinfo{year}{2022}).

\bibitem{hu2021acoustic}
\bibinfo{author}{Hu, G.}, \bibinfo{author}{Tang, L.}, \bibinfo{author}{Liang,
  J.}, \bibinfo{author}{Lan, C.} \& \bibinfo{author}{Das, R.}
\newblock \bibinfo{title}{Acoustic-elastic metamaterials and phononic crystals
  for energy harvesting: A review}.
\newblock \emph{\bibinfo{journal}{Smart Materials and Structures}}
  \textbf{\bibinfo{volume}{30}}, \bibinfo{pages}{085025}
  (\bibinfo{year}{2021}).

\bibitem{charara2025spatially}
\bibinfo{author}{Charara, M.}, \bibinfo{author}{Kujala, Z.},
  \bibinfo{author}{Lee, S.} \& \bibinfo{author}{Gonella, S.}
\newblock \bibinfo{title}{Spatially selective drop-motion programming using
  metamaterials}.
\newblock \emph{\bibinfo{journal}{Proceedings of the Royal Society A}}
  \textbf{\bibinfo{volume}{481}}, \bibinfo{pages}{20240429}
  (\bibinfo{year}{2025}).

\end{thebibliography}


\begin{thebibliography}{12}
\providecommand{\natexlab}[1]{#1}
\providecommand{\url}[1]{\texttt{#1}}
\expandafter\ifx\csname urlstyle\endcsname\relax
  \providecommand{\doi}[1]{doi: #1}\else
  \providecommand{\doi}{doi: \begingroup \urlstyle{rm}\Url}\fi

\bibitem[Telgen et~al.(2024)Telgen, Kannan, Bail, Dorn, Niese, and
  Kochmann]{telgen2024rainbow}
Bastian Telgen, Vignesh Kannan, Jean-Charles Bail, Charles Dorn, Hannah Niese,
  and Dennis~M Kochmann.
\newblock Rainbow trapping of out-of-plane mechanical waves in spatially
  variant beam lattices.
\newblock \emph{Journal of the Mechanics and Physics of Solids}, 191:\penalty0
  105762, 2024.

\bibitem[Dorn and Kochmann(2023{\natexlab{a}})]{dorn2023inverse}
Charles Dorn and Dennis~M Kochmann.
\newblock Inverse design of graded phononic materials via ray tracing.
\newblock \emph{Journal of Applied Physics}, 134\penalty0 (19),
  2023{\natexlab{a}}.

\bibitem[Dorn and Kochmann(2023{\natexlab{b}})]{dorn2023conformally}
Charles Dorn and Dennis~M Kochmann.
\newblock Conformally graded metamaterials for elastic wave guidance.
\newblock \emph{Extreme Mechanics Letters}, 65:\penalty0 102091,
  2023{\natexlab{b}}.

\bibitem[Dorn and Kochmann(2022)]{dorn2022ray}
Charles Dorn and Dennis~M Kochmann.
\newblock Ray theory for elastic wave propagation in graded metamaterials.
\newblock \emph{Journal of the Mechanics and Physics of Solids}, 168:\penalty0
  105049, 2022.

\bibitem[Teh et~al.(2022)Teh, O'Toole, and Gkioulekas]{teh2022adjoint}
Arjun Teh, Matthew O'Toole, and Ioannis Gkioulekas.
\newblock Adjoint nonlinear ray tracing.
\newblock \emph{ACM Transactions on Graphics}, 41\penalty0 (4):\penalty0 1--13,
  2022.

\bibitem[Liu and Nocedal(1989)]{liu1989limited}
Dong~C Liu and Jorge Nocedal.
\newblock On the limited memory {BFGS} method for large scale optimization.
\newblock \emph{Mathematical Programming}, 45\penalty0 (1):\penalty0 503--528,
  1989.

\bibitem[Johnson(2007)]{NLopt}
Steven~G. Johnson.
\newblock The {NLopt} nonlinear-optimization package.
\newblock \url{https://github.com/stevengj/nlopt}, 2007.

\bibitem[Mechanics and Lab(2020)]{ae108}
Mechanics and Materials Lab.
\newblock ae108, 2020.

\bibitem[Hopcroft et~al.(2010)Hopcroft, Nix, and Kenny]{hopcroft2010young}
Matthew~A Hopcroft, William~D Nix, and Thomas~W Kenny.
\newblock What is the {Young's} modulus of silicon?
\newblock \emph{Journal of Microelectromechanical Systems}, 19\penalty0
  (2):\penalty0 229--238, 2010.

\bibitem[Phani et~al.(2006)Phani, Woodhouse, and Fleck]{phani2006wave}
A~Srikantha Phani, J~Woodhouse, and NA~Fleck.
\newblock Wave propagation in two-dimensional periodic lattices.
\newblock \emph{The Journal of the Acoustical Society of America}, 119\penalty0
  (4):\penalty0 1995--2005, 2006.

\bibitem[Zelhofer and Kochmann(2017)]{zelhofer2017acoustic}
Alex~J Zelhofer and Dennis~M Kochmann.
\newblock On acoustic wave beaming in two-dimensional structural lattices.
\newblock \emph{International Journal of Solids and Structures}, 115:\penalty0
  248--269, 2017.

\bibitem[sci()]{scipyButterx2014}
butter 2014; {S}ci{P}y v1.15.2 {M}anual --- docs.scipy.org.
\newblock
  \url{https://docs.scipy.org/doc/scipy/reference/generated/scipy.signal.butter.html}.
\newblock [Accessed 17-04-2025].

\end{thebibliography}

\end{document}


\Large \centering Supplementary information for: ``\textbf{Graded phononic metamaterials: Scalable design meets scalable microfabrication}''\\
\normalsize
by Charles Dorn, Vignesh Kannan, Ute Drechsler, and Dennis M. Kochmann
\begin{justify}

\section{Inverse design}

We provide a detailed description of the inverse design methodology used to generate the designs presented in Figs.~1 and 2 of the main text. Our procedure for the inverse design is composed of the following three steps. First, a unit cell design space is defined and dispersion relations are computed throughout the design space, which is discussed in Section~\ref{sec:defining}. Second, tiles with spatially graded arrangements of unit cells are designed by optimizing ray trajectories, for which the mathematical foundation and numerical implementation are discussed in Sections~\ref{sec:framework} and~\ref{sec:tiledesign}, respectively. Finally, the third step is to assemble the tiles into a desired wave guiding configuration, which is discussed in Section~\ref{sec:assembly}.

\subsection{Defining a design space}
\label{sec:defining}
The first step is to define a parameterized design space of unit cells, which determines the variables that can be spatially graded in a metamaterial consisting of many unit cells. In this work, we define the unit cell design space shown in Fig.~1a of the main text, consisting of elastic beams along each edge, square lattice vectors, and a single geometric design variable $\theta$. This is one of many possible choices for a unit cell design space, but it was carefully selected for the following reasons:

\begin{itemize}
\item We consider beam-based unit cells, for which beam finite elements provide an efficient and physically realistic model (for a comparison of beam elements, solid elements, and experiments in the context of dispersion relations see \cite{telgen2024rainbow}). Furthermore, planar beam architectures exhibit decoupling between in- and out-of-plane modes for the lowest dispersion surfaces \cite{telgen2024rainbow}. Since it is experimentally feasible to excite and measure out-of-plane motion in the wafer samples presented here, this decoupling is advantageous. 

\item For computational efficiency, we define the unit cell design space to have a single parameter (though the approach is sufficiently general to extend to arbitrarily many design parameters). Since the dispersion relation and its first and second derivatives must be numerically evaluated, defining a single parameter design space maximizes efficiency for optimizing complex spatial gradings.

\item A well-chosen unit cell design parameter must be such that its variation leads to a significant change in the dispersion relations. This is essential for enabling wave guiding in graded architectures --- and is the case for $\theta$ in our study. For example, the dispersion surfaces in Fig.~1a of the main text differ by a factor of nearly 4 in their peak value at the extremes of the design space. This allows for substantially different local dispersion relations in different spatial locations in a graded metamaterial. In contrast, if the dispersion surface hardly changes throughout the design space, then the spatial grading would have little effect on how waves propagate and limit the functionality of grading.

\item The unit cell should have four-fold symmetry. This enables the tiles (e.g., Fig.~1b of the main text) to be rotated by $90^\circ$ and mirrored when placed into tile assemblies (e.g., Fig.~2a and d of the main text) without introducing sharp interfaces in unit cell geometry between adjacent tiles.
\end{itemize}
Considering these four factors, the choice of the unit cell design space in Fig.~1a of the main text is justified, though other architectures can also be considered within our design framework. We also note that, throughout this paper, we consider the lowest out-of-plane dispersion surface only to simplify modeling and experiments, but the presented methodology can readily be applied to higher dispersion branches since ray theory is valid at high frequencies \cite{dorn2023inverse,dorn2023conformally}.

Throughout the simulations, physical units are used that match the experimental configuration. The unit cell height and width is $L=100\, \mu$m, the beam thickness is $10\, \mu$m, and the beam width is $5\, \mu$m. The lowest dispersion surfaces computed using beam finite elements at the two extremes of the design space are plotted on the first Brillouin zone in Fig.~1a of the main text; details of the finite element (FE) model used to compute the dispersion relations are discussed in Section~\ref{sec:FEA_bloch}.

\subsection{Theoretical framework for inverse design}
\label{sec:framework}

The tile design step involves optimization of a ray tracing model of wave propagation. Here, ray tracing is used for forward modeling of how waves propagate through spatially graded metamaterials. We refer to \cite{dorn2022ray} for a detailed derivation and description of ray tracing in graded metamaterials. We build an inverse design framework around ray tracing by extending the formulation of \cite{dorn2023inverse} to handle more general cost functions suitable for tile design. 

The starting point is forward modeling of wave propagation in a planar metamaterial using ray tracing. Given the \textit{local} dispersion relation $\omega(\bfk,\bfx)$, which relates the wave vector $\bfk \in \mathbb{R}^2$ to the frequency $\omega\in\Rset$, at a given position $\bfx \in \mathbb{R}^2$ (e.g., computed by Bloch wave analysis for the unit cell at position $\bfx$), wave propagation is described by the ray trajectories satisfying
\begin{align}
   \label{eq:ray1} \dot{\bfx} &= \frac{\pl \omega}{\pl \bfk},\\
   \label{eq:ray2} \dot{\bfk} &= -\frac{\pl \omega}{\pl \bfx}
\end{align}
This constitutes the \textit{ray tracing system}: a system of first-order differential equations that is solved, given initial conditions $\bfx_0 = \bfx(t=0)$ and $\bfk_0=\bfk(t=0)$. Eq.~\eqref{eq:ray1} states that the the group velocity $\bfV = \frac{\pl \omega}{\pl \bfk}$ is tangent to ray path $\bfx(t)$. Eq.~\eqref{eq:ray2} states that changes in wave vector $\bfk$ along the ray path are driven by the spatial change in the local dispersion relation. The ray tracing system can be solved for many rays to efficiently capture where a wave propagates in a graded metamaterial. 

Leveraging ray tracing as an efficient modeling tool, we formulate an optimization problem to design the spatial distribution of unit cells to shape the ray trajectories. Specifically, consider a ray $r$ emerging from an initial position $\bfx_{0r}$ with initial wave vector $\bfk_{0r}$. A contour $\Gamma$ is drawn to enclose $\bfx_{0r}$ that we define as the \textit{exit curve}, which is fixed (see Figure~\ref{fig:S_optimization_setup_general}). Denote $\bfx^*_r$ and $\bfk^*_r$, respectively, as the position and wave vector at which ray $r$ first crosses the exit curve, which occurs at time $t^*_r$. We aim to force the ray to cross through the exit curve at a specified target point $\hat{\bfx}_r$ with a target wave vector $\hat{\bfk}_r$. To achieve the prescribed exit conditions, we minimize the cost function
\begin{align} \label{eq:raycost}
c_r(\bfx^*_r(\bftheta),\bfk^*_r(\bftheta)) = w_1||\bfx_r^*(\bftheta)-\hat{\bfx}_r||^2+w_2||\bfk_r^*(\bftheta)-\hat{\bfk}_r||^2,
\end{align}
where $\bftheta \in \mathbb{R}^N$ is a vector capturing the spatial distribution of design variables, i.e., $\bftheta$ is the vector collecting the design variables at all points within the design domain  (see Section~\ref{sec:tiledesign} for details on defining $\bftheta$). Constants $w_1,w_2\geq 0$ weight each cost function term. Figure~\ref{fig:S_optimization_setup_general} schematically illustrates the variables in the cost function.
We define a constrained optimization problem to minimize $c_r$ of $n_r$ rays, which takes the form
\begin{equation} \label{eq:optimization}
\begin{aligned}
    \min_{\bftheta} C(\bftheta) = &\sum_{r=1}^{n_r} c_r(\bfx^*_r(\bftheta),\bfk^*_r(\bftheta))&&\\
    \text{subject to}\quad & \dot{\bfx}_r = \frac{\partial \omega}{\partial \bfk},  \quad &&t\in[0,t^*_r], \ \ r=1,..., n_r,
    \\ 
    &\dot{\bfk}_r=-\frac{\partial \omega}{\partial \bfx},  \quad &&t\in[0,t^*_r], \ \ r=1,..., n_r,
    \\
    & \calG_r(\bfx^*_r) = 0, \quad &&r=1,..., n_r.
\end{aligned}
\end{equation}
Here, the ray tracing system of Eqs.~\eqref{eq:ray1} and \eqref{eq:ray2} is treated as constraints. Since the exit time $t^*_r$ is not fixed, we impose the additional constraint $\calG_r(\bfx^*_r) = 0$. The quantity $\calG_r$ is defined as the signed distance of the ray at time $t^*_r$ to the exit curve $\Gamma$, so that $\calG_r(\bfx^*_r) = 0$ enforces $t^*_r$ to be the time at which the ray exits $\Gamma$ (see \cite{dorn2023inverse} and \cite{teh2022adjoint} for further discussion).

\begin{figure*}[t]
    \centering
    \includegraphics[width=0.8\linewidth]{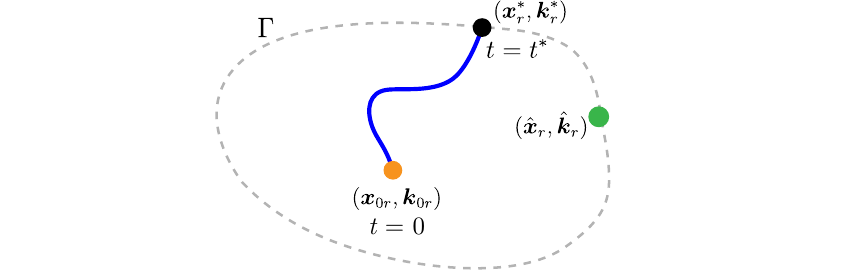}
    \caption{Definitions for setting up the cost function. Ray $r$ has initial conditions $(\bfx_{0r},\bfk_{0r})$ and passes through the exit curve $\Gamma$ at time $t^*$ with state $(\bfx^*_r,\bfk^*_r)$. The optimization seeks to align the exit state of the ray with the desired exit state $(\hat{\bfx}_r,\hat{\bfk}_r)$.}
    \label{fig:S_optimization_setup_general}
\end{figure*}

The optimization problem of Eq.~\eqref{eq:optimization} is solved using gradient-based optimization, for which the total derivative of the cost function with respect to the design variables, i.e., $\frac{\dd C}{\dd \bftheta}$ is required. To efficiently compute this quantity, the adjoint state method is used. Consider the Lagrangian corresponding to the cost function of a single ray (with the subscript $r$ omitted for clarity),
\begin{align}
    \calL(\bftheta,\bfx,\bfk,\bflambda,\bfmu,\rho,t^*) = C-\int_0^{t^*}\left[\bflambda^{\T}\left(\dot{\bfx}-\frac{\pl \omega}{\pl \bfk}\right) + \bfmu^{\T}  \left(\dot{\bfk}+\frac{\pl \omega}{\pl \bfx}\right) \right] \dd t - \rho \calG,
\end{align}
where $\bflambda$, $\bfmu$, and $\rho$ are the Lagrange multipliers corresponding to the three constraint sets of Eq.~\eqref{eq:optimization}, respectively. 

The adjoint state method proceeds by enforcing stationarity of the Lagrangian with respect to all variables except $\bftheta$. Setting the variations with respect to the Lagrange multipliers $\bflambda$, $\bfmu$, and $\rho$ to zero directly return the constraints of Eq. \eqref{eq:optimization}. Setting variations with respect to the state variables $\bfx$ and $\bfk$ to zero returns the adjoint system, which takes the form
\begin{align} 
    \label{eq:adjoint_system1} & \dot{\bflambda} = -\frac{\partial^2 \omega}{\partial \bfx \partial \bfk}\bflambda + \frac{\partial^2 \omega}{\partial \bfx \partial \bfx} \bfmu,
    \\
    \label{eq:adjoint_system2} & \dot{\bfmu} = -\frac{\partial^2 \omega}{\partial \bfk \partial \bfk} \bflambda + \frac{\partial^2 \omega}{\partial \bfk \partial \bfx}\bfmu,
\end{align}
with end conditions
\begin{align}
  \label{eq:lambda_end}&\bflambda(t^*)=\frac{\partial C}{\partial \bfx^*}-\rho \frac{\partial \mathcal{G}}{\partial \bfx^*},\\
  \label{eq:mu_end}
  &\bfmu(t^*)=\frac{\pl C}{\pl \bfk^*}.
\end{align}
Eqs.~\eqref{eq:adjoint_system1}, \eqref{eq:adjoint_system2}, and \eqref{eq:lambda_end} are identical to the adjoint system in \cite{dorn2023inverse}. Here, the end condition for $\bfmu$ in Eq.~\eqref{eq:mu_end} is nonzero, because we consider cost functions dependent on $\bfk$ (which is not the case in \cite{dorn2023inverse}). This is essential in our tile design, as it allows us to control the wave vector with which rays enter and/or leave a tile.

An equation for $\rho$ is obtained by enforcing stationarity of the Lagrangian with respect to $t^*$,
\begin{align} \label{eq:rhodef}
    \rho = \frac{\frac{\pl C}{\pl \bfx^*}\cdot \bfv^* - \frac{\pl C}{\pl \bfk^*} \cdot \frac{\pl \omega}{\pl \bfx^*}}{\frac{\pl \calG}{\pl \bfx^*}\cdot \bfv^*}.
\end{align}
The form of $\rho$ in Eq. \eqref{eq:rhodef} differs from that in \cite{dorn2023inverse}, again because here we consider a cost function dependent on $\bfk^*$. 

Finally, since variations of the Lagrangian with respect to all variables except $\bftheta$ are zero, the gradient of the cost function is easily obtained from
\begin{align} \label{eq:gradient_final}
    \frac{\dd C}{\dd \bftheta} = \frac{\delta \calL}{\delta \bftheta} =  -\int_0^{t^*} \left( -\bflambda^{\T} \frac{\partial^2 \omega}{\partial \bfk \partial \bftheta} + \bfmu^{\T} \frac{\partial^2 \omega}{\partial \bfx \partial \bftheta} \right) \dd t,
\end{align}
where all quantities in the integrand are known after forward and reverse ray tracing.

The formulations of this section generally follow those of \cite{dorn2023inverse}, which provides more detailed derivations and explanations. However, we here consider two distinct differences to generalize the framework and make it suitable for tile design. First, an arbitrary exit curve is considered here in contrast to a strictly circular exit curve in \cite{dorn2023inverse}. Second, the cost function defined in Eq.~\eqref{eq:raycost} is a function of both $\bfx^*$ and $\bfk^*$, while \cite{dorn2023inverse} considers cost functions only dependent on $\bfx^*$. Consequently, Eqs.~\eqref{eq:mu_end} and \eqref{eq:rhodef} differ from the analogous equations (19) and (12) of \cite{dorn2023inverse} due to contributions from $\bfk^*$-dependence of the cost function $C$.

\subsection{Numerical implementation of tile design}
\label{sec:tiledesign}

Using the optimization framework of Section~\ref{sec:framework}, we present the design of two tiles, which are shown in Fig.~1c of the main text. In this case, the optimization aims to design the spatial distribution of the single unit cell design variable $\theta$ to satisfy two different objective functions that are outlined in Sections~\ref{sec:tile1design} and~\ref{sec:tile2design}.

Since ray tracing solutions conserve frequency (as long as the local dispersion relation is time-independent, which is the case for linear elastic materials), each ray has a fixed frequency. We consider optimization of many rays at the same fixed frequency $\omega_0$, which we take as $\omega_0=750$ kHz throughout all examples. One could consider rays of multiple frequencies in the objective function, as was done in \cite{dorn2023inverse}, but we observed that our designs exhibited sufficiently broad frequency range surrounding the design frequency to demonstrate experimentally. Section~\ref{sec:FEA_transient} discusses further the behavior at different frequencies.

For both tiles, a vector $\bftheta \in \mathbb{R}^N$ is defined to parameterize the design problem. Entries of $\bftheta$ correspond to values of $\theta$ on a regular $n\times n$ square grid, such that $N=n^2$. We consider a design grid that spans 64$\times$64 unit cells, corresponding to the size of the tile. The spacing of the grid of design variables may or may not align with the physical lattice of unit cells. This allows for control over the overall number of design variables. If the design variable grid is taken to have the same spacing as the unit cells, then each design variable corresponds to one unit cell's design. Alternatively, a coarser grid of design variables can be chosen, from which the design of a specific unit cell can be interpolated. Control over the coarseness of the design grid allows for adaptive spatial refinement during optimization, which is beneficial for finding spatially smooth optimal solutions \cite{dorn2023inverse,teh2022adjoint}. We begin the optimization with a design grid spacing of $16L$ and gradually refine it to a spacing of $L$ (such that each unit cell has its own entry in $\bftheta$) during the optimization.

Following the formulations of Section~\ref{sec:framework}, a three-step procedure is followed to compute the cost function gradient. First, given a set of ray initial conditions and design, the ray is traced by solving Eqs.~\eqref{eq:ray1}-\eqref{eq:ray2}. Second, the adjoint system of Eqs.~\eqref{eq:adjoint_system1}-\eqref{eq:adjoint_system2} is solved, starting from the end of the ray and marching backwards in time. Finally, the cost function gradient is computed by evaluating the integral of Eq.~\eqref{eq:gradient_final} along each ray, which is given to an optimization algorithm to update the design variables. A fourth-order Runge-Kutta solver is used to solve both the ray tracing and adjoint systems, while trapezoidal integration along the ray path is used to evaluate Eq.~\eqref{eq:gradient_final}. The computed cost function gradient is used to inform the L-BFGS algorithm \cite{liu1989limited} for gradient-based optimization, for which we use the NLopt implementation \cite{NLopt} for all examples. 

A key challenge in this optimization is that we seek solutions that are not only global minima, but which send the cost function of Eq.~\eqref{eq:optimization} to zero such that the desired exit conditions are satisfied for each ray. This is critical for tile assembly, since any misalignment in the ray trajectories at the exit of a tile results in a misaligned incident wave to the adjacent tile, thus corrupting the wave guiding capabilities in assemblies of tiles. In the presented examples, there was sufficient design freedom and smoothness to find solutions that zero the cost function, using gradient-based optimization. Future work could explore global optimization in pursuit of tile designs with additional functionalities.

\subsubsection{Tile 1: Point source to plane wave}
\label{sec:tile1design}

The objective of Tile 1 is to convert a point excitation at its center to plane waves exiting the four tile edges (i.e., rays exiting perpendicular to the tile edges). For this example, we take $n_r=100$ rays in in the cost function. The initial conditions of each ray are determined by the point excitation at the center, such that $\bfx_{0r}=\bfzero$ for all rays. The set of all possible wave vector initial conditions corresponding to the excitation frequency lie on the isofrequency contour of the local dispersion surface at the excitation point, i.e., $\omega(\bfk,\theta_0)=\omega_0$ where $\theta_0=\theta(\bfx_0)$. The direction that the ray exits the excitation point is the direction of the vector $\bfV(\bfk_{0r},\bfx_{0r})$ (i.e., the direction of the initial group velocity). We choose values of $\bfk_{0r}$ such that the rays uniformly span initial angles from $0$ to $45^\circ$ and enforce eight-fold symmetry in the distribution of $\theta$ over the tile, such that the rays between $0$ and $45^\circ$ can be rotated/reflected to capture all rays exiting the point source.

The unit cell design parameter $\theta$ is set to a fixed value $\theta_0$ at the point $\bfx_0$ to ensure that the local dispersion relation at the excitation point remains constant throughout the optimization, so that each ray's initial conditions remain constant as well. Additionally, $\theta$ is fixed at $\theta_0$ on the boundary of the tiles (at $x=\pm32L$ and $y=\pm32L$) as well as everywhere outside of the tile, which ensures continuity in $\theta$ between adjacent tiles during the tile assembly step. In this study, we found through numerical experiments that $\theta_0=0.4$ is a suitable value.

To achieve the design of Tile 1, the exit contour is defined as a semicircle to the left of $x=64L$ with a radius of $64L$. In the cost function of Eq.~\eqref{eq:raycost}, only the exit conditions $\hat{x}=0$ and $\hat{k}_y=0$ are included. The $y$-coordinate of the exit point is left free, and $k_x$ at the exit does not need to be prescribed due to the four-fold symmetry of the dispersion relations (enforcing $k_y=0$ is sufficient for ensuring the group velocity is horizontal). The corresponding cost function seeks rays that exit the plane at $x=32L$ horizontally. Finally, eight-fold symmetry in the $\theta$-distribution is enforced about $x=0$, $y=0$, $x=y$, and $x=-y$. These symmetries are implemented by mirroring the cost function gradient about these planes to directly enforce symmetry before handing it to the optimizer). The resulting ray trajectories can be mirrored about the symmetry axes, as plotted in Fig.~1c of the main text.

\subsubsection{Tile 2: Plane wave $90^\circ$ turn}
\label{sec:tile2design}

The objective of Tile 2 is to steer a plane wave incident to the left boundary of the tile to a plane wave exiting out the bottom of the tile. We consider $n_r = 80$ rays in the cost function.  Around the perimeter of the tile, the unit cell design is fixed at $\theta_0=0.4L$ to ensure compatibility with other tiles. Rays are considered with an initial $x_0=-32L$ and uniformly spaced $y$-coordinates spanning $y=\pm 20L$ (for a $64L\times 64L$ tile centered at the origin). The initial wave vector is the same for all rays, which is obtained by taking the point on the isofrequency contour $\omega(\bfk,\theta_0)=\omega_0$ corresponding to a group velocity $\bfV(\bfk_0, \theta_0)$ that is horizontal.

In the cost function, the target exit wave vector is $\hat{k}_x=0$ with $\hat{k}_y$ left unspecified, which is sufficient for enforcing that the rays exit vertically due to the four-fold symmetry of the dispersion relations. To ensure the rays exit out the bottom of the tile, we take $\hat{y}=-32L$. The value of $\hat{x}$ is defined independently for each ray as $\hat{x}=y_0$, which was carefully chosen to allow a solution mimicking an interface that deflects rays downward, as observed in the plot of Fig.~1c in the main text. We found this to be a reliable approach to finding solutions for turning a plane wave by $90^\circ$. In the resulting design, the interface is diffusely spread over several unit cells due to the spatial grading, so that the ray trajectories remain smooth and a sharp interface is avoided. The exit curve is taken as a semicircle of radius $64L$ centered at the bottom left corner of the tile. We enforce symmetry of the $\theta$-distribution about $x=y$, which we found to aid in converging to a solution.

\subsection{Tile assembly}
\label{sec:assembly}

Since each tile is designed to be compatible with an equal and constant value of $\theta=0.4L$ around all boundaries, their rays can be connected to construct a continuous solution when the two tiles are placed next to each other. This is because the ray tracing equations depend only on the local dispersion relations, which we have ensured are continuous across tile boundaries. Thus, taking these tiles as building blocks can lead to many customized tile assemblies that guide waves in different ways. We note that since the dispersion relations exhibit four-fold symmetry about the $k_x=0$ and $k_y=0$ axes (following the four-fold symmetry of the unit cell geometry), the tiles can be reflected and rotated during their assembly while ensuring their ray solutions follow the same reflections/rotations. Two examples of tile assemblies are presented in Fig.~2a and 2d of the main text. Both designs are validated by transient finite element simulations (see Section \ref{sec:FEA}), and the design of Fig.~2a is also demonstrated experimentally. An additional assembled design example is shown in Fig.~\ref{fig:S_assembly_3}, which demonstrates scalability of the modular tile-based design approach to achieve a more complex guided wave path. This design spans approximately 143,000 unit cells to guide waves emerging from either instance of Tile 1 along a complicated path corresponding to the rays of Fig.~\ref{fig:S_assembly_3}b.

\begin{figure*}[t]
    \centering
    \includegraphics[width=\linewidth]{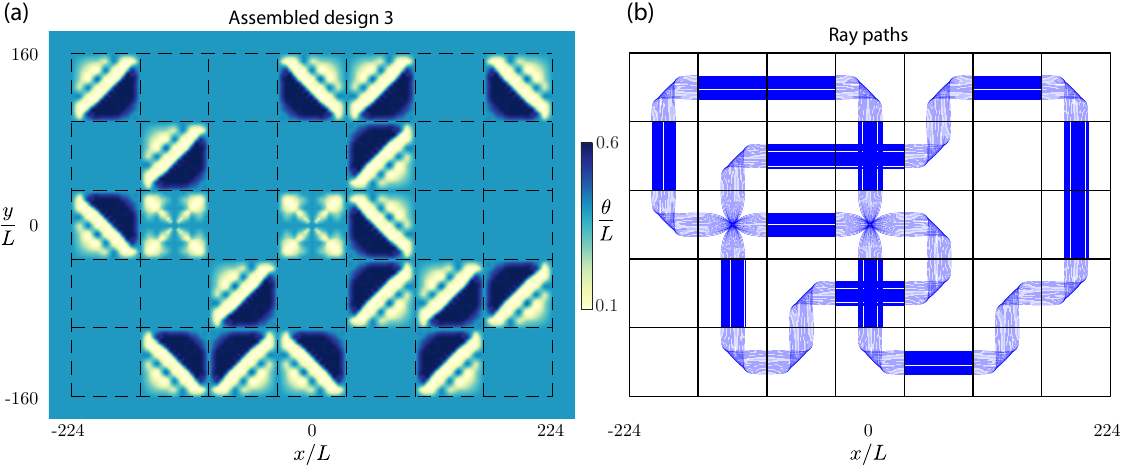}
    \caption{Demonstration of a complex tile assembly design spanning approximately 143,000 unit cells. (a) Spatial distribution of the unit cell design variable $\theta$, with dotted lines marking tile boundaries. (b) The corresponding ray trajectories.}
    \label{fig:S_assembly_3}
\end{figure*}

\section{Finite element modeling}
\label{sec:FEA}

We use an FE model of beam-based metamaterials to perform two types of computations. The first is the computation of dispersion relations for unit cells throughout the design space, which is the initial step in the numerical inverse design process. The second use is transient dynamic simulations on a finite domain containing the assembled design to validate its wave guiding capability and enable quantitative comparison to experiment.

For both computations, Timoshenko beam finite elements are used to discretize the metamaterial, which is a reliable and efficient approach for physically realistic modeling of wave propagation in beam lattices \cite{telgen2024rainbow}. The open-source C++ finite element code \textit{ae108} \cite{ae108} is used. We use meshes with a maximum element size of $L/10$, which was found to be sufficiently dense to accurately capture the lowest out-of-plane mode. The orthotropic anisotropy of silicon is accounted for by using the directional moduli determined by the in-plane orientation of each beam. In the prototype and the FE model, the lattice vectors are aligned with the $\bar{1}10$ and $110$ directions of the silicon crystal structure; see \cite{hopcroft2010young} for the wafer orientation definitions and corresponding orthotropic moduli, which were adopted here.

\subsection{Dispersion relation computations}
\label{sec:FEA_bloch}

Dispersion surfaces are computed based on a finite element model of a unit cell. For a given unit cell, the computational procedure involves first applying Bloch boundary conditions to the unit cell and solving the resulting eigenvalue(s) for a given wave vector $\bfk$; see, e.g., \cite{phani2006wave}. Dispersion surfaces are constructed by solving the Bloch eigenvalue problem for values of $\bfk$ spanning the first Brillouin zone. For planar unit cells modeled with beam elements, in- and out-of-planes are decoupled \cite{zelhofer2017acoustic}. That is, out-of-plane excitation will only excite out-of-plane motion. Since we experimentally measure out-of-plane motion, we limit our interest to dispersion surfaces corresponding to out-of-plane mode shapes. Without loss of generality, we only consider the lowest out-of-plane dispersion surface (agreement between simulation results and our experiments confirm that this is indeed sufficient).

The inverse design framework requires the dispersion relation $\omega(k_1, k_2, \theta)$ and its first and second partial derivatives to be evaluated for a given wave vector and $\theta$-value. To obtain the dispersion surface throughout the unit cell design space ($0.1L\leq \theta \leq 0.6L$), we solve the Bloch eigenvalue problem on a $50\times50$ grid in $k$-space, spanning the first Brillouin zone for 100 values of $\theta$ ranging from $0.1L$ to $0.6L$. The dispersion surfaces corresponding to $\theta=0.1L$ and $\theta=0.6L$ are shown in Fig.~1a of the main text. Finite differences are used to approximate the first and second derivatives of the dispersion surface with respect to $\bfk$ and $\theta$, which are also needed for optimization. Evaluations of the dispersion surface and its derivatives for any given values of $\bfk$ and $\theta$ are then obtained by interpolation.

\subsection{Transient dynamic simulations}
\label{sec:FEA_transient}

Transient finite element simulations are performed to numerically validate the designed tile assemblies of Fig.~2 of the main text, as well as for comparison to experiments. For simulations of both designs, a finite domain of 300$\times$300 unit cells (3 cm $\times$ 3 cm) is modeled. To match the experimental setting, a pulse displacement excitation is applied in the out-of-plane direction at the junction of the unit cell at the center of the center tile (at $x=y=0$). The pulse is a half-sine wave of duration $1 \ \mu s$. While not exactly representative of the broadband experimental pulse excitation, this idealized excitation spans approximately the same frequency range. The outer boundary of the simulation domain is clamped and the entire domain is initially at rest. A Newmark-beta integration scheme (with parameters $\gamma=0.5$ and $\beta=0.25$, thus avoiding numerical damping) is used to solve for the dynamic response with a time step of $0.22\, \mu$s.
\begin{figure*}[ht]
    \centering
    \includegraphics[width=\linewidth]{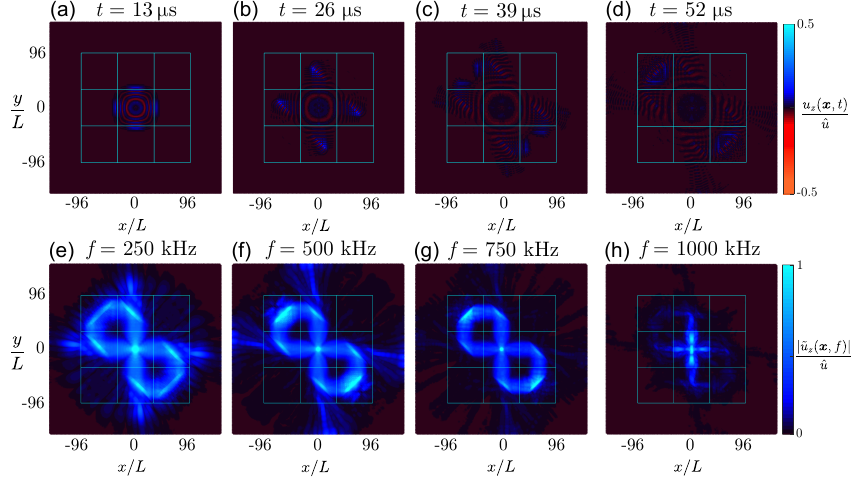}
    \caption{Snapshots of the wave field in the transient dynamic finite element simulation of the figure-eight design at (a) 13~$\mu$s, (b) 26~$\mu$s, (c) 39~$\mu$s, and (d) 52~$\mu$s. The amplitude of the temporal FFT of the displacement signal at each spatial location is plotted at (e) 250~kHz, (f) 500~kHz, (g) 750~kHz (the design frequency), and (h) 1000~kHz.}
    \label{fig:FEA_snapshots_figure8}
\end{figure*}

For the figure-eight design of Fig.~2a of the main text, the response is simulated for a duration of $55 \, \mu$s, which is sufficient time for the wave to complete the figure-eight trajectory. Snapshots of the resulting wave field are plotted in Fig.~\ref{fig:FEA_snapshots_figure8}a-d at times $t=$~13, 26, 39, and 52 $\mu$s, where the color map corresponds to the instantaneous out-of-plane displacement. The simulation results show strong wave guiding despite the broadband excitation. Note that, as no dissipative boundary conditions are imposed (all boundaries are clamped) and the dynamic solver suppresses numerical damping, any observed wave attenuation and redirection effects stem from the spatially graded design.

Although the spatial grading is designed specifically for wave guiding at 750 kHz, the operating bandwidth spans a much wider range than the immediate vicinity of the design frequency. This is evident in Fig. \ref{fig:FEA_snapshots_figure8}e-h, which shows the amplitude of the frequency spectrum of the out-of-plane displacement, denoted $\tilde{u}_z(\bfx,f)$, at frequency snapshots $f=$~250, 500, 750, and 1000~kHz. While the cleanest wave guiding is observed at the design frequency of 750 kHz, strong wave guiding is observed down to 250 kHz.

For the cross design of Fig.~2d of the main text, the response is simulated for a duration of $104\, \mu$s to allow sufficient time for the wave to travel along the entire design trajectory. Snapshots of the resulting wave field are plotted in Fig.~\ref{fig:FEA_snapshots_cross}a-d at times $t=$~13, 52, 78, and 104~$\mu$s, where the color map corresponds to the instantaneous out-of-plane displacement. Again, wave guiding is observed over a broad frequency spectrum, as shown in the snapshots of the frequency spectrum in Fig.~\ref{fig:FEA_snapshots_cross}.

For clear visualization of wave guiding, we also performed transient simulations with harmonic excitation at the design frequency $f=750$~kHz. The results are shown in Supplementary Videos 1 and 2.

\begin{figure*}[ht]
    \centering
    \includegraphics[width=\linewidth]{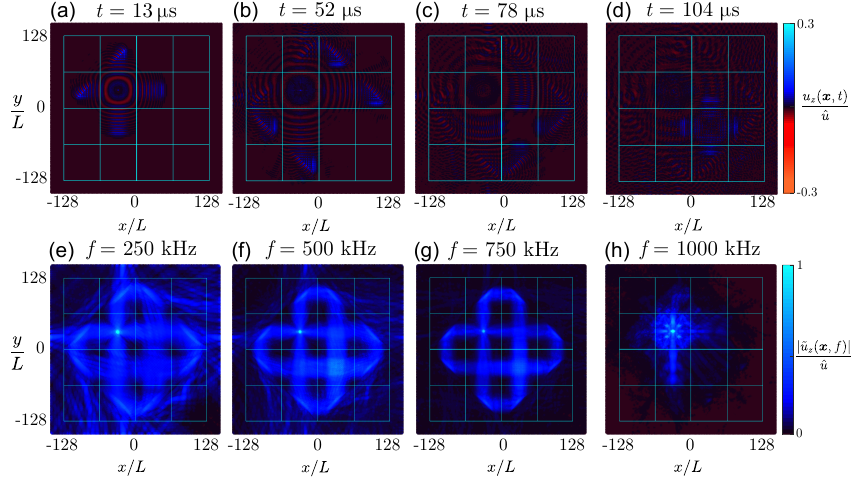}
    \caption{Snapshots of the wave field in the transient dynamic finite element simulation of the cross design at (a) 13, (b) 52, (c) 78, and (d) 104~$\mu$s. The amplitude of the temporal FFT of the displacement signal at each spatial location is plotted at (e) 250, (f) 500, (g) 750 (the design frequency), and (h) 1000~kHz.}
    \label{fig:FEA_snapshots_cross}
\end{figure*}

\section{Microfabrication}

Samples were fabricated using commercially-procured 100~mm diameter Silicon-On-Insulator (SOI) wafers with three layers: a single-crystal silicon device layer (thickness~$10 \pm 0.5$~\textmu m), a buried oxide (BOX) layer (thickness~$2.4 \pm 0.12$~\textmu m), and a single-crystal silicon handle layer (thickness~$400 \pm 10$~\textmu m). The fabrication protocol involved etching windows in the handle layer, the designed architecture in the device layer, and release of the architected device layer by etching away the intermediate BOX layer. The following sequence of steps was followed during fabrication. 
\begin{itemize}[leftmargin=*]
    \item \textbf{Back window etching}:~The first sequence of steps was developed to etch windows into the device layer for pump excitation of the acoustic wave (see Methods section in the main paper). Before this sequence, a thin layer ($\sim 3$~\textmu m) of sacrificial positive photoresist (AZ4533) was deposited and baked (120$^\circ$C for 2~minutes) on the device layer to protect this surface from damage or contamination. A photomask was custom designed and written by a direct laser writer (DLW), commonly used for optical lithography, at the BRNC Cleanroom facility in IBM Zurich. The handle layer was spin-coated with positive photoresist AZ4562, exposed using the photomask, and developed by a standard photolithography protocol (developer:~dilute KOH solution (AZ400K, 1:3 ratio) for $\approx$ 40~seconds), followed by Deep Reactive Ion Etching~(DRIE) for 60-90~minutes. 
    \item \textbf{Device etching}:~The next sequence of steps involved etching the architecture in the device layer, again using a high-resolution custom-developed photomask. A home-built Python code generates the input files for photomask writing, using the computational design output files. This infrastructure allows for the seamless handover between computational design and fabrication, which may also be automated. The same standard photolithography techniques were used as in the previous step, albeit using a different photoresist (AZ4533), hence a shorter time for development (25~seconds). This was followed by DRIE for $\sim 15$~minutes.
    \item \textbf{Photoresist stripping}:~Photoresist layers were carefully stripped by first rinsing with acetone and isopropyl alcohol, followed by immersion in dimethyl sulfoxide (DMSO) at 120$^\circ$C for $\sim10$~minutes. 
    \item \textbf{Release and post-fabrication treatment}:~The intermediate BOX layer was removed using dry HF etching to generate free-standing architected films of $10$~\textmu m thickness. As a final step, $20 - 50$~nm thin aluminium films were deposited on each side of the sample to minimize the penetration of lasers during acoustic pump-probe measurements (see the Methods section of the main text for details).
\end{itemize}

\section{Data analysis}
Standard Fourier analysis was performed on the raw time series data (averaged over 50 samples) at each measurement point. Due to the broadband excitation of the experiment relative to the computational design, each data set was filtered using a Butterworth bandpass filter \cite{scipyButterx2014} of order 15 --- the choice of filter order was to minimize the loss of signal at the edges of the pass band. For the data presented in Fig.~4 of the main text, a pass band of width $1.6$~MHz centered at $1.4$~MHz was chosen. 

Broadband excitation can provide further insight into the frequency-dependent wave-guiding response of the sample. To this end, data was further filtered using narrower pass bands with multiple center frequencies $\omega_c$ between $400$~kHz and $1.5$~MHz, and width $\Delta\omega=200$~kHz. Results of this analysis and comparison with FE simulation data are discussed in Sec.~\ref{appsec:freqwave}. FE simulation data were analyzed using the same parameters and numerical framework as the experimental data. 

\section{Frequency-dependent wave guiding:~experiments vs.\ simulations} \label{appsec:freqwave}

\begin{figure*}[ht]
    \centering
    \includegraphics[width=\linewidth]{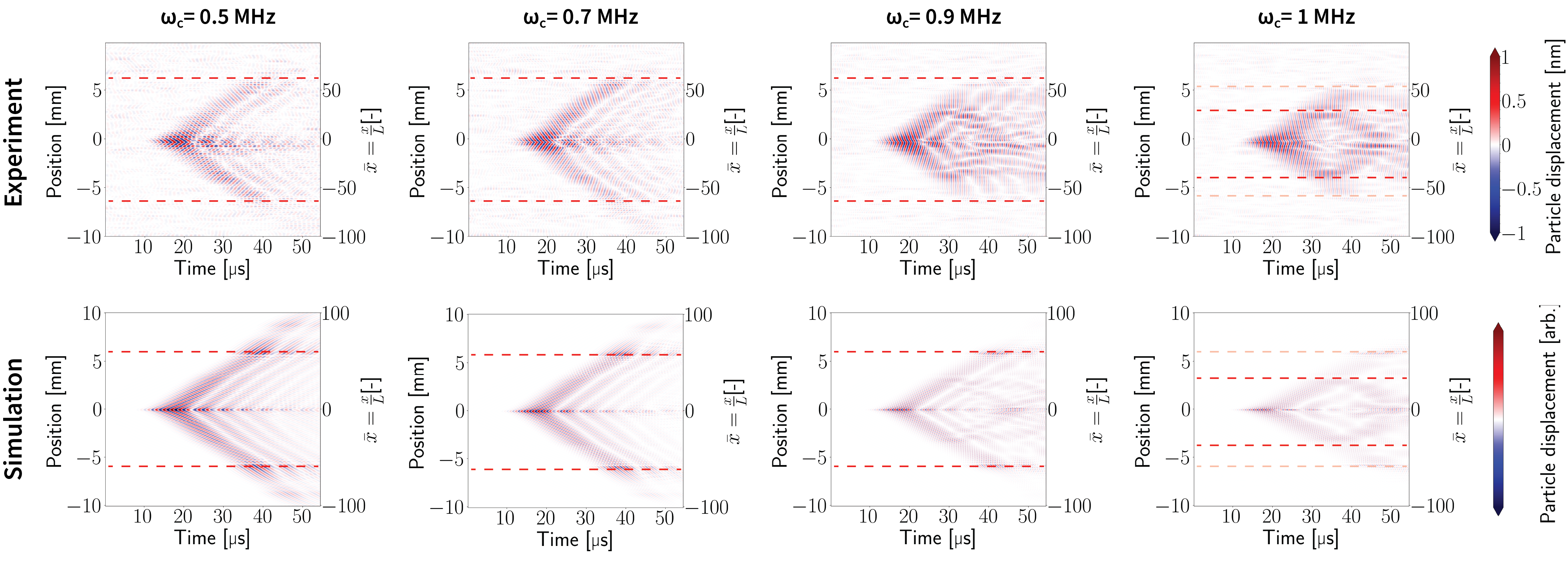}
    \caption{Bandpass-filtered results from line scan L1 (filter width $\Delta \omega = 200$~kHz): comparison between experiment and simulation shows excellent agreement as well as wave guidance over a broad frequency range. The red dashed lines indicate the positions beyond which no wave propagation is observed. Above 0.8~MHz, back-reflection effects are noticeable in both experiments and simulations, which emerge within the figure-8 and shall not be confused with reflections from the outer boundaries.}
    \label{fig:freq_analysis_L1}
\end{figure*}

Figs.~\ref{fig:freq_analysis_L1}, \ref{fig:freq_analysis_L2}, and \ref{fig:freq_analysis_L3} show results from the frequency-dependent analysis described in the previous section along lines L1, L2, and L3, respectively, as defined in the main text. Each top row shows data from experiments, while the corresponding bottom row shows equivalent FE simulation data. Each column corresponds to a specific center frequency (a constant filter width of $200$~kHz was maintained). In general, we observe excellent agreement between experimental and simulated data. Note that the computational inverse design was for a target frequency of 750~kHz. 

\begin{figure*}[!h]
    \centering
    \includegraphics[width=\linewidth]{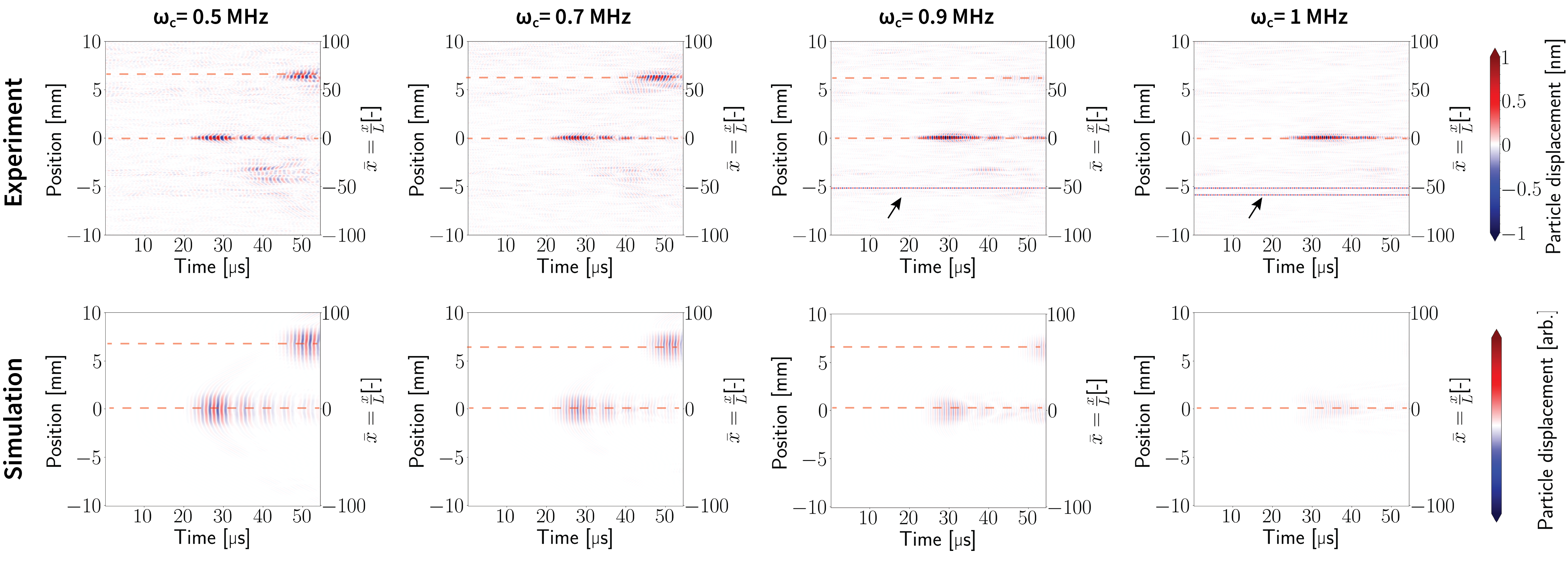}
    \caption{Bandpass-filtered results from line scan L2 (filter width $\Delta \omega = 200$~kHz): comparison between experiments and simulations. The red dashed lines indicate the positions beyond which no wave propagation is observed. No significant signal is observed beyond the excitation point at center frequencies $\omega_c>0.8$~MHz. Noisy data at two measurement positions, to be discarded, are marked by black arrows.}
    \label{fig:freq_analysis_L2}
\end{figure*}

Line scans L2 (Fig.~\ref{fig:freq_analysis_L2}) do not show significant differences from the broadband data (Fig.~4 of the main manuscript) at the center frequencies $\omega_c=0.5$~MHz and $0.7$~MHz. At 0.9~MHz and 1~MHz, however, we notice that there is almost no visible signal beyond the excitation point. This is consistent with the observations from line scan L1 (Fig.~\ref{fig:freq_analysis_L1}): wave modes at higher frequencies (close to 1~MHz) seem to be ``trapped'' within line L1 and the horizontal line connecting the centers of L1 and L2. Hence, high-frequency waves are trapped between two perpendicular lines intersecting at the excitation point, with waves being attenuated before the reflection points expected from the figure-eight wave guiding response. 

\begin{figure*}[!h]
    \centering
    \includegraphics[width=\linewidth]{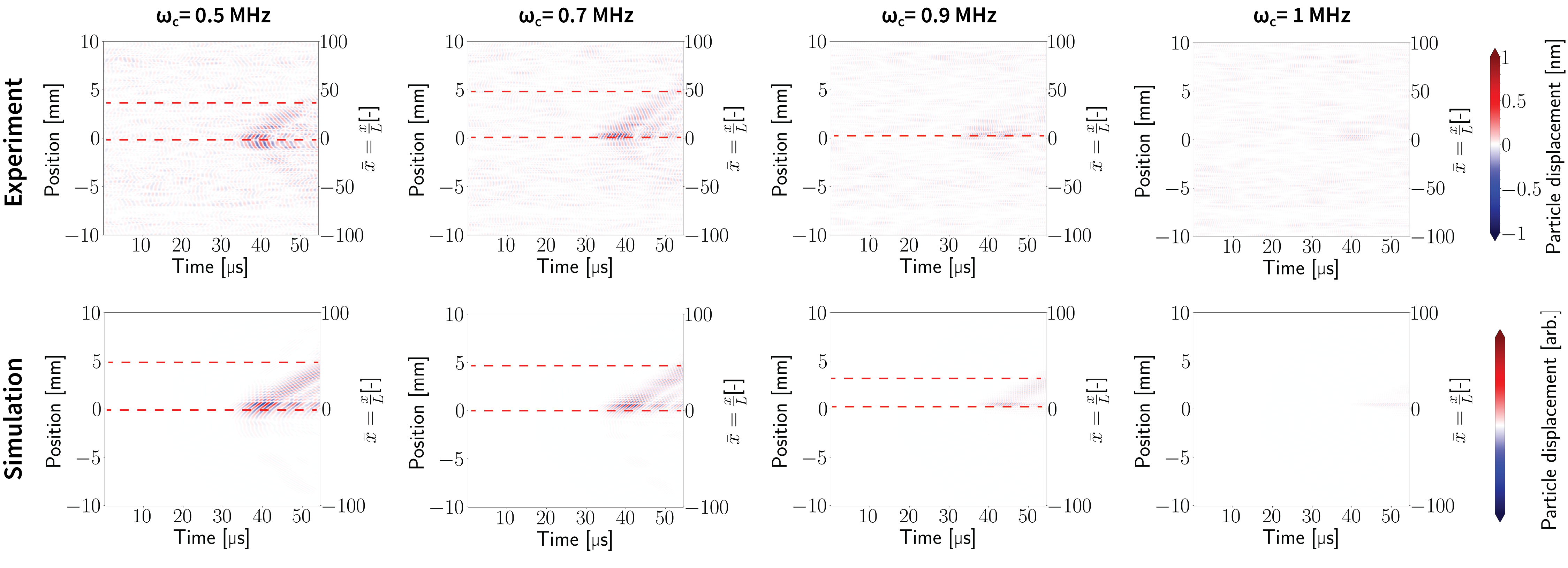}
    \caption{Bandpass-filtered results from line scan L3 (filter width $\Delta \omega = 200$~kHz): comparison between experiments and simulations. The red dashed lines indicate the positions beyond which no wave propagation is observed. No significant signal is observed beyond the excitation point at center frequencies beyond $\omega_c>0.8$~MHz. This confirms wave trapping along the horizontal $x$-axis at those frequencies.}
    \label{fig:freq_analysis_L3}
\end{figure*} 

This observation is supported by data from line scan L3 (Fig.~\ref{fig:freq_analysis_L3}), showing that this wave has attenuated significantly by the time it reaches the center of line~L3. The latter is apparent from the reduction in signal amplitude at center frequencies of 0.9 and 1~MHz in Fig.~\ref{fig:freq_analysis_L3}. While this could in principle be an effect of other dissipation mechanisms (e.g., air damping), these effects were much less pronounced at lower frequencies, which hints at a structural, dispersive origin of the observed high-frequency wave attenuation through our graded architecture. 
\end{justify}

\newpage
\bibliography{bibliography}